\documentclass[12pt]{article}
\usepackage{amsmath,amsfonts,amsthm,amssymb}
\usepackage{slashed}
\usepackage{xcolor}
\usepackage{graphicx}
\usepackage{epstopdf}
\usepackage{array,booktabs}
\usepackage{hyperref}
\usepackage{tabulary}
\usepackage{multirow}
\usepackage{array,arydshln}
\usepackage{bbm} 
\usepackage{cite}
\usepackage{mathtools} 

\hypersetup{linktoc = all}  
\hypersetup{pdfborderstyle={/S/U/W 0.5}}

\numberwithin{equation}{section}
\newcommand{\beq}{\begin{equation}}
\newcommand{\eeq}{\end{equation}}

\newcommand{\address}[1]{\vbox{\center\em#1}}
\renewcommand{\title}[1]{\vbox{\center\LARGE{#1}}\vspace{5mm}}

\DeclareMathOperator{\Tr}{Tr}

\newcommand{\rf}[1]{\eqref{#1}}
\declareslashed{}{/}{-.08}{0}{V} 

\makeatletter
\newcommand*{\letterdef@}{}
\newcommand*{\letterdef}[3]{%
  \def\letterdef@##1{\expandafter\newcommand\csname #1\endcsname{#2{##1}}}%
  \@tfor\@tempa :=#3\do{\expandafter\letterdef@\expandafter{\@tempa}}}
\makeatother
\letterdef{bb#1}{\mathbb} {CHPRZ}    
\letterdef{c#1}{\mathcal}{ABCDEFGHIJKLMNOPQRSTUVWXYZ} 
\letterdef{rm#1}{\mathrm} {dDF} 

\newcommand{\bs}{\bar{\sigma}}

\begin{document}
\bibliographystyle{utphys}


\begin{titlepage}

\hfill{}
\\
\vspace{15mm}

\center{{\LARGE Gauge Theory Dynamics and    K\"ahler Potential  \\[18pt]
  for Calabi-Yau Complex Moduli}}
\vspace{10mm}

\begin{center}
\renewcommand{\thefootnote}{$\alph{footnote}$}
Nima Doroud,\footnote{\href{mailto:ndoroud@perimeterinstitute.ca}{\tt ndoroud@perimeterinstitute.ca}}
Jaume Gomis\,\footnote{\href{mailto:jgomis@perimeterinstitute.ca}{\tt jgomis@perimeterinstitute.ca}}

\vspace{10mm}

\address{${}^{a,b}$Perimeter Institute for Theoretical Physics,\\
Waterloo, Ontario, N2L 2Y5, Canada}
\address{${}^{a}$Department of Physics, University of Waterloo,\\
Waterloo, Ontario N2L 3G1, Canada}

\renewcommand{\thefootnote}{\arabic{footnote}}
\setcounter{footnote}{0}

\end{center}

\vspace{25mm}

\abstract{
\medskip\medskip
\normalsize{
\noindent
 We compute the  exact two-sphere  partition function and matrix of two-point functions of  operators in the chiral ring with their complex conjugates   in two-dimensional supersymmetric gauge theories. For  gauge theories that flow in the infrared to a Calabi-Yau nonlinear sigma model, these renormalization group invariant observables determine 
  the exact K\"ahler potential and associated Zamolodchikov metric in the complex structure moduli space of the Calabi-Yau manifold.
}
}

\vspace{10mm}

\noindent
\vfill

\end{titlepage}

\tableofcontents


\newpage

\section{Introduction}
\label{sec:intro}

String theory compactifications  on a Calabi-Yau manifold accommodate physically  appealing models of particle physics within an ultraviolet complete theory of quantum gravity. 
In string theory, there is a  beautiful and rich interplay between the four dimensional effective field theory capturing the low energy dynamics of string theory on a Calabi-Yau threefold and the two dimensional 
superconformal field  theory (SCFT) on the string worldsheet. Massless scalar fields in four dimensions  correspond to exactly marginal operators in the SCFT realizing  the Calabi-Yau 
compactification, while  the couplings encoding the effective field theory dynamics are captured by worldsheet correlation functions.

Metric   deformations of a  Calabi-Yau manifold give rise to massless four dimensional scalar fields.  They split into two families: K\"ahler class deformations  and complex structure deformations. 
The number of  corresponding moduli  is  determined by the two non-trivial Betti numbers of a Calabi-Yau threefold: $h^{1,1}$ and $h^{1,2}$. In the worldsheet theory, the exactly marginal operators associated 
to K\"ahler class  and complex structure moduli are  in different     multiplets
\begin{align}
  \cO_{i} &\Longleftrightarrow   \text{K\"ahler moduli} \ \ \ \ \   i=1, \ldots, h^{1,1}\nonumber \\
  \cO_{a} &\Longleftrightarrow \text{complex   moduli} \ \ \ a=1, \ldots, h^{1,2}\,\,. \nonumber
\end{align}
The operators $\cO_{i}$ and $ \cO_{a}$ are  primary
 operators annihilated by different supercharges $(q_A$ vs. $q_B)$ and $U(1)_R\times U(1)_\cA$ R-symmetry generators in the $\cN=(2,2)$ superconformal algebra  
\begin{align}
[q_A, \cO_{i}  ]&=[R,\cO_{i} ]=0\label{annihilateA}\\[3pt]
 [q_B, \cO_{a}]&=[{\cal A}, \cO_{a} ]=0\,.
\label{annihilateB}
\end{align}
These operators form a ring in the  SCFT: $\cO_{i}$ the twisted chiral ring and $\cO_{a}$ the chiral ring.\footnote{These operators also play an important role in massive $\cN=(2,2)$ field theories.} 
The conjugate operators  
$\cO_{\bar i}$ and $\cO_{\bar a}$       are annihilated by   conjugate   supercharges $q_{\bar A}$ and $q_{\bar B}$.

Upon compactification on a Calabi-Yau, the metric in field space for the moduli in four dimensions is non-trivial. It is  given by the metric in the ``quantum"
K\"ahler moduli space $\cM_K$ and the complex structure moduli space  $\cM_C$ respectively.  In the worldsheet theory,  the moduli metric is captured by the Zamolodchikov metric  of the CFT:  the two-point function of the 
corresponding exactly marginal operators on the two-sphere.
The metric in the (complexified)   K\"ahler moduli space  is\footnote{Superconformal Ward identities determine the  two-point functions of the exactly marginal operators   in terms of those for the primary operators 
$\cO_{i}$ and $ \cO_{a}$.}
\beq
  G^{K}_{i\bar j}=\langle \cO_{i} (N)\,   \cO_{\bar j}(S)\rangle_{S^2} \,,
\label{metricA}
\eeq
while the metric in the complex structure moduli space is
\beq
  G^{C}_{a\bar b}=\langle \cO _{a}(N)\,     \cO_{\bar b}(S)\rangle_{S^2}\,.
\label{metricB}
\eeq
Operators are inserted at the north pole of the two-sphere and conjugate operators at the south pole. Both metrics are K\"ahler, and are 
  determined by the K\"ahler potential $ \cK^K$ and $\cK^C$ in the   K\"ahler moduli space $\cM_K$ and   complex structure moduli space  $\cM_C$  
\beq
  G^{K}_{i\bar j}=\partial_i\partial_{\bar j} \cK^K\qquad\qquad  G^{C}_{a\bar b}=\partial_a\partial_{\bar b} \cK^C\,.
\eeq
These, in turn, also determine the superpotential (Yukawa) couplings for the four dimensional chiral and antichiral multiplets appearing in heterotic string compactifications.  

 The computation of these observables in a Calabi-Yau nonlinear sigma model (NLSM) is notoriously difficult.
 A fruitful approach towards  NLSMs is to  study instead two dimensional gauge theories \cite{Witten:1993yc} -- known as gauged linear sigma models  (GLSM's) -- which flow in the infrared to a NLSM. In favorable situations,  
 renormalization group invariant observables in the  ultraviolet GLSM  can be computed exactly by supersymmetric localization \cite{Witten:1988ze, Witten:1991zz, Pestun:2007rz} of functional integrals. In this optimal scenario,  
 computations in the GLSM  capture exactly correlators in the elusive infrared  NLSM.

Recently,  the   computation of two dimensional gauge theories   in \cite{Doroud:2012xw} (see also \cite{Benini:2012ui})  has resulted in    the {\it exact}   K\"ahler potential $\cK_{K}$ in the 
``quantum" K\"ahler moduli space $\cM_K$ of a  Calabi-Yau manifold in terms of the GLSM  two-sphere partition function 
\beq
  \cZ_{A}= e^{-\cK_{K}}\,,
\eeq
as   conjectured in \cite{Jockers:2012dk} and demonstrated  in \cite{Gomis:2012wy}. This provides a physics-based approach to computing, in particular, worldsheet instanton corrections to the K\"ahler potential  
that does not rely on mirror symmetry \cite{Dixon:1987bg,Lerche:1989uy,Candelas:1990rm}. This result has found a variety of applications and generalizations. Most notably the computation of novel Gromov-Witten invariants   
\cite{Jockers:2012dk} for which no other method of computation is currently available as well as new insights into mirror symmetry \cite{Gomis:2012wy} and   D-branes \cite{Sugishita:2013jca,Honda:2013uca,Hori:2013ika}.

 In this paper we identify and compute the   observable which captures the exact K\"ahler potential $\cK_{C}$ in the complex structure moduli space $\cM_C$ of a large class Calabi-Yau manifolds. It is given by    
 a different GLSM  two-sphere partition function   
\beq
  \cZ_{B}= e^{-\cK_{C}}\,,
  \label{complex}
\eeq
which  now depends on the complex structure moduli. In this setup, the two-sphere   can be enriched by inserting in a supersymmetric way chiral operators $\cO_a$ at the north pole and their conjugates  
$\cO_{\bar b}$  at the south pole, while in  \cite{Doroud:2012xw,Benini:2012ui} the admissible supersymmetric insertions were twisted chiral operators $\cO_i$ and their conjugates $\cO_{\bar j}$ at the poles.

Our results yield a purely gauge theory realization of the K\"ahler potential $\cK_{C}$, which admits a geometrical representation in terms of the nowhere vanishing top holomorphic form  $\Omega$ of a Calabi-Yau manifold as
\beq
e^{-\cK_{C}}=i^{\,\text{dim}M}\int_M \Omega\wedge \overline \Omega\,.
\label{formulacplex}
\eeq
In this paper we focus on gauge theories with an abelian gauge group, which can describe Calabi-Yau manifolds which are complete intersections in toric varieties, leaving the results for non-abelian gauge theories to a separate publication \cite{Gomis:2013}.

 The plan of the rest of the paper is as follows. In section \ref{sec:susyglsm} we explain how gauge theories on a two-sphere can   realize two different supersymmetry algebras, and how the choice of supersymmetry
 determines which operators can be added in the functional integral in a supersymmetric way. This, in turn,  determines which supersymmetry algebra has to be realized in a GLSM in order to compute the  
 Zamolodchikov metric for K\"ahler moduli and for complex structure moduli. In section \ref{sec:twistedmultiplets} we write down the supersymmetry transformations and  action of two-dimensional gauge theories which 
 allow the insertion of operators in the chiral ring (and their conjugates). In section \ref{sec:localization} we compute the exact two-sphere partition function and matrix of two-point functions of abelian gauge 
 theories by supersymmetric localization.  In section \ref{sec:CYG} we apply our results to the computation of the two-sphere partition of GLSM for various families of Calabi-Yau geometries, and obtain a gauge 
 theory realization of the exact K\"ahler potential in the complex structure moduli space
 \rf{formulacplex}. Section \ref{sec:disc} contains some discussion. Technical computations and details have been delegated to the appendices.

\section{Supersymmetry on \texorpdfstring{$S^2$}{S2}  And Gauged Linear Sigma Models}
\label{sec:susyglsm}

The nonlinear sigma model  describing a Calabi-Yau compactification is a two dimensional $\cN=(2,2)$ SCFT. 
Generally, it is only at special loci in the SCFT moduli space --   spanned by the twisted chiral and chiral operator marginal couplings --  that explicit computations can be performed. 
This moduli space acquires a beautiful geometric interpretation as the ``quantum" K\"ahler and complex structure moduli space of the Calabi-Yau geometry. A window  into the dynamics of such a SCFT throughout 
moduli space   can be   provided by
  a   gauged linear sigma model \cite {Witten:1993yc}, a super-renormalizable gauge theory  which flows to the Calabi-Yau  NLSM in the infrared.

The corresponding Zamolodchikov metric on the SCFT moduli spaces \rf{metricB} and \rf{metricA} is the matrix of two-point functions on the two-sphere of chiral and twisted chiral operators with their conjugates.
The goal is to realize  these as supersymmetric correlators in the ultraviolet GLSM where exact computations can be performed using supersymmetric localization, and infer from them the results 
for the infrared SCFT. Computing these observables in the   GLSM requires constructing $\cN=(2,2)$ gauge theories on $S^2$. 

While Calabi-Yau sigma models on the two-sphere have  $\cN=(2,2)$ superconformal symmetry,\footnote{Classically, a NLSM can be placed on the two-sphere by a Weyl transformation while preserving 
the full $\cN=(2,2)$ superconformal algebra. For a NLSM with a Calabi-Yau target space, the superconformal symmetry remains quantum mechanically.} a gauge theory   on $S^2$ can realize at most   an $SU(2|1)$ 
subalgebra of the  superconformal algebra. 
An $SU(2|1)$ subalgebra  is generated by supercharges   in the $\cN=(2,2)$  superconformal algebra that close into the $SU(2)$ isometry generators $J_m$ of the two-sphere together with $T$, 
one of the  $U(1)_R\times U(1)_\cA$ $R$-symmetry generators in the superconformal algebra.  
It is given by
\beq
\begin{aligned}
 \ [J_{m},J_{n}] &= i \epsilon_{mnp} J_p~
& [J_{m},Q_\alpha] &= -\frac{1}{2} \gamma_{m~ \alpha}^{~\,\beta} Q_\beta\qquad
  &  [J_{m},S_ \alpha] &=-\frac{1}{2} \gamma_{m~ \alpha}^{~\,\beta}S_\beta
\\
\{S_ \alpha,Q_\beta\} &= \gamma^{m}_{\alpha \beta} J_{m} - \frac{1}{2} C_ {\alpha\beta}T~ &
[T,Q_\alpha] &= -Q_\alpha  \qquad &
  [T,S_\alpha] &=S_\alpha\,.
\end{aligned}
\label{SU(2|1)}
\eeq

There are two inequivalent $SU(2|1)$ subalgebras on the two-sphere:  $SU(2|1)_A$ and  $SU(2|1)_B$. These two algebras are mapped into each other by the $\mathbb{Z}_2$ mirror (outer) automorphism $\sigma$ 
of the $\cN=(2,2)$ superconformal algebra\footnote{See appendix \ref{app:susy} for the embedding of these two algebras in the $\cN=(2,2)$ superconformal algebra on $S^2$  \cite{Doroud:2012xw}  and the action of the mirror 
automorphism on the generators.} \cite{Lerche:1989uy}
\begin{equation}
\label{outer}
\begin{aligned}
  &\hspace{-12pt}SU(2|1)_A \\
  &Q_A	\\
  &S_A	\\
  &R	  \\
  &J_m
\end{aligned}
\hspace{50pt}
  \xLeftrightarrow{\ \ \ \ \sigma \ \ \ \ }
\hspace{65pt}
\begin{aligned}
  &\hspace{-12pt}SU(2|1)_B\\
  &Q_B\\
  &S_B\\
  &\cA\\
  & J_m
\end{aligned}
\end{equation}

The choice of supersymmetry  realized in the gauge theory  determines which observables of the theory in the infrared   can  be  captured by localizing correlators in the ultraviolet gauge theory.
The $SU(2|1)_A$ (resp. $SU(2|1)_B)$  algebra has  a supercharge $\cQ_A$ (resp. $\cQ_B$) that interpolates between  $q_A$ (resp. $q_B$) at the north pole and   $q_{\bar A}$ (resp. $q_{\bar B}$) 
at the south pole of the two-sphere.
In $SU(2|1)_A$ the conserved R-symmetry is $T=R$ while in $SU(2|1)_B$ the conserved R-symmetry is $T=\cA$. This implies that $\cQ_A$  in $SU(2|1)_A$  annihilates  twisted chiral   operators  $\cO_i$  
at the north pole and    their conjugates $\cO_{\bar i}$ at south  pole \rf{annihilateA}, while    $\cQ_B$ in $SU(2|1)_B$     annihilates      chiral  operators $\cO_a$ 
at the north pole    and their conjugates $\cO_{\bar a}$ at the south  pole \rf{annihilateB}.  As a result, the correlators in the GLSM that flow to the Zamolodchikov metric in the infrared SCFT are supersymmetric.

The $SU(2|1)_A$  and $SU(2|1)_B$ invariant gauge theory two-sphere partition function and two-point functions  just described  are independent of the gauge coupling   
(see section \ref{sec:deco} for  decoupling theorems). Gauge coupling independence implies that these gauge theory observables  are renormalization group invariants. In particular, 
they coincide with those in the theory in the  extreme infrared, where $g_{\text{YM}}^{2}\rightarrow \infty$. In particular, for a Calabi-Yau GLSM, this is
 none other than the sought-after  infrared SCFT. Therefore,  by virtue of the properties of $\cQ_A$ and $\cQ_B$, the computation of the   Zamolodchikov metric on the K\"ahler   (resp. complex structure) moduli space 
 can be  recast in terms of    correlators in an    $SU(2|1)_A$ (resp. $SU(2|1)_B$) invariant GLSM  on $S^2$.  
 \medskip
 \begin{center}
\begin{tabular}{c|c|c|c|c}
  & North Pole & South Pole & Correlation Function & Metric  
  \\
  \hline
  $\cQ_{A}$ & $\begin{array}{c}
		q_{A} \\[3pt] \cO_{i}\\ 
              \end{array}$
              &
              $\begin{array}{c}
		q_{\bar{A}} \\[3pt] \cO_{\bar{j}} 
	      \end{array}$
	      &
	      $\langle \cO_{i}\, \cO_{\bar{j}}\rangle$
	      &
	     K\"{a}hler moduli  
	        \\[5pt]
  \hline
  $\cQ_{B}$ & $\begin{array}{c}
		q_{B} \\[3pt] \cO_{a} 
              \end{array}$
              &
              $\begin{array}{c}
		q_{\bar{B}} \\[3pt] \cO_{\bar{b}}  
	      \end{array}$
	      &
	      $\langle \cO_{a}\, \cO_{\bar{b}}\rangle$
	      &
	     Complex moduli\end{tabular}
\end{center}

\subsection{Gauged and Nonlinear Sigma Models on $S^2$}

 The multiplets with which to construct $SU(2|1)_A$ and $SU(2|1)_B$-invariant theories are those of two dimensional $\cN=(2,2)$ supersymmetry.
 In flat spacetime, a chiral   superfield $\Phi$
and twisted chiral superfield $Y$ \cite{Gates:1984nk}  
obey\footnote{The superspace derivatives are $D_{\pm}={\frac{\partial}{\partial \theta^\pm}} -i \bar\theta^\pm \partial_\pm$ and $\bar D_{\pm}=-{\frac{\partial}{\partial \bar\theta^\pm}} 
+i \theta^\pm \partial_\pm$.} 
\beq
\bar D_+ \Phi =\bar D_- \Phi=0\,,\qquad\qquad \bar D_+ Y = D_- Y=0\,.
\eeq
A vector multiplet is a   superfield $V=V^\dagger$ subject to   gauge redundancy   by a chiral multiplet $\Phi$
\beq
  V\simeq V+i\left(\Phi- \bar \Phi\right)\,,
\eeq
while  for a twisted vector multiplet the gauge redundancy is by a twisted chiral multiplet $Y$  
\beq
 V\simeq V+i\left(Y- \bar Y\right)\,.
\label{tv}
\eeq
Conventionally, a NLSM on a K\"ahler manifold is written in terms of $\cN=(2,2)$ chiral multiplets. 
Since supersymmetric minimal coupling between chiral multiplets and twisted vector multiplets do not exist, an ultraviolet GLSM flowing to   a NLSM is  
described by an $\cN=(2,2)$ vector multiplet with gauge group $G$ coupled to a chiral multiplet in a (reducible) representation ${\bf R}$ of $G$.

On chiral multiplets the $U(1)$ R-symmetry $R$ acts vectorially while the $U(1)$ axial R-symmetry $\cA$ acts axially. Therefore 
a gauge theory  with an R-invariant superpotential  can be placed on $S^2$ while preserving $SU(2|1)_A$. 
 The current associated to $\cA$ however, can be anomalous and only gauge theories for which the anomaly cancels can be placed on the two-sphere
while preserving $SU(2|1)_B$. If  a   gauge theory in flat space preserves both $R$ and $\cA$, it is expected to flow in the infrared to an $\cN=(2,2)$ SCFT. 
Such gauge theories are the basis for the construction of Calabi-Yau GLSM's.

In   a Calabi-Yau GLSM based on vector and chiral multiplets,   the complex structure moduli of the Calabi-Yau are realized by background expectation values for chiral multiplets while the K\"ahler moduli
are realized by background expectation values for twisted chiral multiplets.\footnote{The FI and topological couplings can be written in terms of  the field strength multiplet which is a twisted chiral superfield.}  
These moduli appear in the superpotential $\cW$ and twisted superpotential $W$ of the GLSM,
which are $\cQ_A$-exact and $\cQ_B$-exact respectively (see section \ref{sec:deco}):
 \medskip
\begin{center}
\begin{tabular}{|c|c|}
\hline
  $\cW$  & $W$   \\
  \hline
  $\cQ_{A}$- exact & $\cQ_{B}$-exact \\
  \hline
   \end{tabular}
\end{center}
\vskip+10pt
Therefore, in an  $SU(2|1)_A$-invariant   GLSM  based on vector and chiral multiplets,  
the partition function depends only on the K\"ahler moduli,  while in an $SU(2|1)_B$-invariant GLSM of vector and chiral multiplets, the partition function depends only on the complex structure 
moduli.\footnote{A Calabi-Yau GLSM may not realize all the complex structure and K\"ahler moduli of the Calabi-Yau.}

In \cite{Doroud:2012xw}, the off-shell  supersymmetry transformations for a vector multiplet and charged chiral multiplet realizing the full $\cN=(2,2)$ superconformal algebra were written down 
(see appendix B in \cite{Doroud:2012xw}).  This includes the supersymmetry transformations for both the  $SU(2|1)_A$ and the $SU(2|1)_B$ subalgebra of the superconformal algebra. Nevertheless, 
in this paper we proceed with the computation of the  $SU(2|1)_B$-invariant partition function using a different route (the $SU(2|1)_A$ computation was done in \cite{Doroud:2012xw,Benini:2012ui}).

 An alternative way 
 to proceed is to change the coordinates used to write down the GLSM (and ensuing infrared NLSM). In this equivalent approach, the  gauge theory realizes the  $SU(2|1)_A$ symmetry but   is written in terms of   
 twisted vector and twisted chiral multiplets. These multiplets are the image of the  vector and chiral multiplets under the action of the 
 mirror automorphism \rf{outer}, which also exchanges $SU(2|1)_B$ with $SU(2|1)_A$. The gauge theory constructed by realizing the $SU(2|1)_{A}$ on the twisted vector and twisted chiral multiplets is equivalent
 to the gauge theory constructed by realizing the $SU(2|1)_{B}$ on the vector and chiral multiplets.
 
In this choice of coordinates, the K\"ahler moduli of the Calabi-Yau  are now background expectation values for chiral multiplets\footnote{The FI and topological couplings can  be written in terms of  
the twisted field strength multiplet,  a  chiral superfield.} while complex structure moduli are  background expectation values for twisted chiral multiplets. These appear in superpotential and twisted 
superpotential terms respectively. Since superpotential terms are $\cQ_A$-exact,    the partition function only depends on complex structure moduli.

 The two approaches described above are  equivalent. If the gauge theory is a Calabi-Yau GLSM, 
 the partition function   computes the K\"ahler potential on the complex structure moduli space.

\section{Complex Structure Gauged Linear Sigma Model}
\label{sec:twistedmultiplets}

The  goal of this section is to construct $SU(2|1)_A$-invariant two-dimensional $\cN=(2,2)$ gauge theories of twisted vector multiplets and 
twisted chiral multiplets. In flat space the Lagrangian of a vector coupled to a chiral multiplet is identical to the Lagrangian of a twisted vector coupled to a twisted chiral multiplet. 
This is no longer the case when the theory is placed on the two-sphere. The background fields \cite{Festuccia:2011ws} and curvature couplings needed to couple the theory to the two-sphere in a supersymmetric way are different, 
and thus the resulting Lagrangians are different. We now proceed to construct the supersymmetry transformations and invariant couplings for the twisted vector and twisted chiral multiplets.

\subsection{Twisted Vector Multiplet}

An $\mathcal{N}=(2,2)$ twisted vector multiplet consists of a real vector, two complex scalars related by complex conjugation, two complex spinors and a real auxiliary scalar 
$(A_\mu, \sigma,\bar\sigma, \eta,\bar \eta, \rmD)$, all of which are valued in the Lie algebra of the gauge group $G$. While a twisted vector multiplet and a vector multiplet with the same gauge group $G$ have exactly 
the same field content, the supersymmetry transformations on the two multiplets are realized differently. 

The $SU(2|1)_A$ supersymmetry transformations on the twisted vector multiplet fields are  
\begin{equation}
\label{tvmsusy}
\begin{aligned}
  \delta \eta &= i \slashed{D}(\sigma \epsilon) + \bar{\epsilon} (\rmD + i F) - \frac{ i }{2} \gamma^{\hat 3} \bar{\epsilon} [\sigma, \bs]
  \\
  \delta \bar{\eta} &= i \slashed{D}(\bs \bar{\epsilon}) + \epsilon (\rmD - i F) - \frac{ i }{2} \gamma^{\hat 3} \epsilon [\sigma, \bs]
\\
\delta{A_{\mu}} &= \frac{ i }{2}(\epsilon\gamma^{\hat 3}\gamma_{\mu} \eta - \bar{\epsilon}\gamma^{\hat 3}\gamma_{\mu}\bar{\eta})
  \\
  \delta \sigma &= \bar{\epsilon} \eta \\
  \delta \bs &= \epsilon \bar{\eta}\\
  \delta \rmD &= \frac{ i }{2}\left\{D_{\mu}(\epsilon \gamma^{\mu} \eta) - [\sigma, \epsilon \gamma^{\hat 3} \bar{\eta}] \right\}
	      + \frac{ i }{2}\left\{D_{\mu}(\bar{\epsilon} \gamma^{\mu} \bar{\eta}) + [\bs, \bar{\epsilon} \gamma^{\hat 3} \eta] \right\}\,.
\end{aligned}
\end{equation}
They are parametrized by conformal Killing spinors $\epsilon$ and $\bar\epsilon$ obeying
\beq
\nabla_\mu \epsilon=\frac{1}{2r} \gamma_\mu \gamma^{\hat 3} \epsilon\qquad \nabla_\mu \bar \epsilon=-\frac{1}{2r} \gamma_\mu \gamma^{\hat 3}\bar \epsilon\,,
\label{killl}
\eeq
where $r$ is the radius of the two-sphere. 
These transformations realize the $SU(2|1)_A$ algebra off-shell up to gauge transformations. Concretely, the resulting algebra is  
\begin{equation}
\label{algebra}
\begin{aligned}
  [\delta_{\epsilon_{1}},\delta_{\epsilon_{2}}] &= \delta_G(\Lambda) 
  \\
  [\delta_{\bar{\epsilon}_{1}},\delta_{\bar{\epsilon}_{2}}] &=\delta_G(\bar \Lambda)  
  \\
  [\,\delta_{\epsilon},\, \delta_{\bar{\epsilon}}\,] &= \delta_{SU(2)}(v)+\delta_R(\alpha)+\delta_G(\Omega) 
\end{aligned}
\end{equation}
where the $SU(2)$ isometry transformation is constructed from the $S^2$ Killing vector
\begin{equation}
\label{spinorbilinears}
v=i\bar{\epsilon}\gamma^{\mu}\epsilon\, \partial_{\mu} \,,
\end{equation}
and the $U(1)_R$ transformation is parametrized by the scalar
\begin{equation}
\label{spinorbilinearsb}
  \alpha = -\frac{1}{2r}\bar{\epsilon}\gamma^{\hat 3}\epsilon \,.
\end{equation}
The $R$-charges of the various fields are:\footnote{These are the $U(1)_\cA$ charges of  the vector multiplet fields for a vector multiplet   of  vanishing   $U(1)_\cA$ charge.}
\medskip
\begin{center}
  \begin{tabular}{*{8}{>{\(}c<{\)}}}
    \toprule
    \sigma & \eta_{+} & \eta_{-} & A_\mu & \rmD & \bar{\eta}_{+} & \bar{\eta}_{-} & \bar{\sigma}
    \\
    -2 & -1 & -1 & 0 & 0 & +1 & +1 & +2 
    \\
    \bottomrule
  \end{tabular}
\end{center}
\medskip
Finally, the field dependent gauge transformation parameters generated in the closure of the algebra are
\beq
\Lambda=- \epsilon_{2}\gamma^{\hat 3}\epsilon_{1}\sigma\qquad\quad \bar\Lambda=  \bar{\epsilon}_{2}\gamma^{\hat 3}\bar{\epsilon}_{1}  \bar{\sigma}\qquad \quad \Omega=- v^{\mu} A_{\mu}\,.
\label{gaugetransf}
\eeq

\subsection{Twisted Chiral Multiplet}

The field content of a twisted chiral multiplet is the same as the standard chiral multiplet but also has different supersymmetry transformations.
A twisted chiral multiplet can be minimally coupled in a supersymmetric way to a twisted vector multiplet. It transforms in a    representation ${\bf R}$ of the gauge group $G$. The $SU(2|1)_A$ supersymmetry 
transformations,    invariant action  and partition function of uncharged 
 twisted chiral multiplets on $S^{2}$ appeared in \cite{Gomis:2012wy}.

 The $SU(2|1)_A$ supersymmetry transformations of   charged  twisted chiral  multiplet fields \newline
 $(Y,\bar Y,\zeta,\bar{\zeta},G,\bar G)$ are
\begin{equation}
\label{tcmsusy}
\begin{aligned}
\delta Y &= (\bar{\epsilon}\gamma_{-}-\epsilon\gamma_{+}) \zeta
  \\[5pt]
  \delta \bar{Y} &= (\bar{\epsilon}\gamma_{+}-\epsilon\gamma_{-}) \bar{\zeta}
  \\[5pt]
  \delta \zeta_{+} &= - \gamma_{+} ( i \slashed{D} Y - G ) \bar{\epsilon} + i \gamma_{+} \epsilon \sigma Y
  \\[5pt]
   \delta \zeta_{-} &= +\gamma_{-} ( i \slashed{D} Y - G ) \epsilon - i \gamma_{-} \bar{\epsilon} \bar{\sigma} Y
  \\[5pt]
\delta \bar{\zeta}_{+} &= +\gamma_{+}( i \slashed{D} \bar{Y} - \bar{G}) \epsilon - i \gamma_{+} \bar{\epsilon} \, \bar{Y} \bar{\sigma} 
  \\[5pt]
  \delta \bar{\zeta}_{-} &= -\gamma_{-}( i \slashed{D} \bar{Y} - \bar{G}) \bar{\epsilon} + i \gamma_{-} \epsilon \, \bar{Y} \sigma 
  \\[5pt]
  \delta G &= + i \epsilon \gamma_{-} \left( \slashed{D} \zeta  - \eta Y - \sigma  \zeta \right)
	  - i \bar{\epsilon} \gamma_{+} \left( \slashed{D} \zeta + \bar{\eta} Y - \bar{\sigma} \zeta \right)  \\[5pt]
  \delta \bar{G} &= + i \epsilon \gamma_{+} \left( \slashed{D} \bar{\zeta} - \bar{Y} \eta - \bar{\zeta} \sigma  \right)
	  - i \bar{\epsilon} \gamma_{-} \left( \slashed{D} \bar{\zeta} + \bar{Y} \bar{\eta} - \bar{\zeta} \bar{\sigma} \right)\,.
\end{aligned}
\end{equation}
These supersymmetry transformations realize the off-shell $SU(2|1)$ algebra \eqref{algebra} with the same parameters and with the following $R$-charge assignments:\footnote{These are the same as the $U(1)_\cA$  
charges of the components of a chiral superfield with vanishing $U(1)_\cA$ charge.}    
\medskip
\begin{center}
  \begin{tabular}{*{8}{>{\(}c<{\)}}}
    \toprule
    \bar{G} & \bar{Y} & \bar{\zeta}_{-} & \bar{\zeta}_{+} & \zeta_{-} & \zeta_{+} & Y & G
    \\
    0 & 0 & -1 & +1 & +1 & -1 & 0 & 0
    \\
    \bottomrule
  \end{tabular}
\end{center}
\medskip
The supersymmetry transformations of a twisted chiral multiplet of $U(1)_\cA$ charge $\Delta$ can be obtained from \rf{tcmsusy} by the  field redefinition  \cite{Gomis:2012wy}
\beq
G\rightarrow G+\frac{\Delta}{2r} Y\,.
\eeq
Since correlators do  not depend on $\Delta$, we take it to vanish.

The $U(1)_{R}$ transformation acts chirally on the twisted chiral multiplet fermions $\zeta$ and $\bar{\zeta}$. Since the $U(1)$ R-symmetry charge $R$ appears explicitly in the anticommutator of supercharges in 
$SU(2|1)_A$, anomaly cancellation of $R$ is required  to write down an  $SU(2|1)_A$ supersymmetric theory of twisted vectors and twisted chirals on the two-sphere. The   $R$-current is quantum mechanically conserved  
whenever the sum of the gauge charges of all charged twisted chiral multiplets vanish for each abelian gauge group factor in $G$. This guarantees that if the  flat space gauge theory is also invariant under the
R-symmetry $\cA$, that the gauge theory   flows in  the infrared to  an $\cN=(2,2)$ SCFT, and if it has a geometrical phase,   to a  Calabi-Yau NLSM.

\subsection{Supersymmetric Lagrangian}
\label{susylag}
We now write down the $SU(2|1)_A$-invariant action for a twisted vector multiplet coupled to a charged twisted chiral multiplet. The action has several couplings that are separately supersymmetric
\begin{equation}
\label{action}
  S= S_{\text{t.v.m.}} + S_{\text{FI}} + S_{\text{top}} + S_{\text{t.c.m.}} + S_{W} + S_{\overline{W}}\,.
\end{equation}
The supersymmetrized kinetic terms for the twisted vector multiplets fields are 
\begin{align}
\label{tvmL}
  \cL_{\text{t.v.m.}} &= \frac{1}{2 g_{\text{YM}}^{2}}\Tr\left\{ F^{2} + D^{\mu}\bar{\sigma}D_{\mu} \sigma + \frac{1}{4} [\sigma, \bar{\sigma}]^{2} + \rmD^{2} - i \bar{\eta} 
      \left( \slashed{D} + \frac{1}{r} \gamma^{\hat 3} \right) \eta + i \bar{\sigma}(\eta\gamma^{\hat 3}\eta) - i \sigma (\bar{\eta}\gamma^{\hat 3}\bar{\eta}) \right\}\,,
 \end{align}
where $F\equiv \frac{1}{2} \epsilon^{\mu\nu} F_{\mu\nu}$.
The supersymmetric Lagrangian for the charged twisted chiral multiplet fields is
\begin{align}
\label{tcmL}
  \cL_{\text{t.c.m.}} &=  \bar{Y} \left(-D_{\mu}^{2} + i \rmD + \frac{ \{\sigma,\bar{\sigma}\} }{2}  \right) Y + \bar{G}G + i \bar{Y} ( \bar{\eta}_{-} - \eta_{+}) \zeta
      + i \bar{\zeta}(\bar{\eta}_{+} - \eta_{-} )Y + i \bar{\zeta} (\slashed{D} -\bar{\sigma} \gamma_{+} -\sigma \gamma_{-})
      \zeta\,.
\end{align}
Twisted chiral multiplets couple via a  twisted superpotential\footnote{We use here a convenient normalization.}   $W$ 
\begin{align}
 \cL_{W} &= {i\over 4\pi}  \left(  W^{\prime \prime}(Y) \zeta_{+} \zeta_{-} - W^{\prime}(Y) G + \frac{i}{r} W(Y)  \right)\,.
 \label{twistedW}
\end{align}
Each $U(1)$ factor in the gauge group  admits a supersymmetric  Fayet-Iliopoulos (FI)  and topological term
\begin{align}
  \cL_{\text{top}} + \cL_{\text{FI}} &= -i \Tr \left( \frac{\vartheta}{2\pi} F + \xi  \rmD \right) \,.
  \label{FItop}
 \end{align}
For each abelian factor, the associated field strength multiplet $\Sigma$ is a chiral superfield, and the FI and topological term  can be encoded in a linear superpotential $\cW$
\beq
\cW=\frac{i\tau}{2} \Sigma\,,
\label{superFI}
\eeq
where
\beq
\cL_\cW=\frac{\partial\cW}{\partial\Sigma} F_{\Sigma} - \frac{\partial^{2}\cW}{\partial \Sigma^{2}}\eta_{+}\eta_{-}\,.
\label{supergeneral}
\eeq
Superpotential couplings are $SU(2|1)_A$ invariant if the superpotential $\cW$ carries $R$-charge $-2$, which is the charge of  $\Sigma$.  
For twisted vector multiplets on $S^2$, $SU(2|1)_A$-invariance implies that the complexified FI parameter  
\beq
\tau=\frac{\vartheta}{2\pi} +i \xi\,
\label{complexFI}
\eeq
is an exactly marginal coupling.

The action in flat space, obtained by sending $r\rightarrow \infty$ in our expressions, has an additional $U(1)_\cA$ R-symmetry if the charge of the twisted superpotential $W$ is $-2$. On the two-sphere, 
however,  the non-minimal $1/r$ couplings in the action required by supersymmetry break this  $U(1)_\cA$ R-symmetry. This breaking can  be understood as arising due to the non-trivial background fields in 
the supergravity multiplet required to couple the gauge theory to a supersymmetric supergravity background \cite{Festuccia:2011ws}.

The parameters of the ultraviolet GLSM are the gauge couplings for each gauge group factor, the complex parameters appearing in the twisted superpotential and the complexified FI parameters appearing in the 
superpotential. We note that unlike $SU(2|1)_A$-invariant GLSM's based on vector and chiral multiplets, the twisted chiral multiplets have vanishing twisted masses, since the scalars in the twisted vector multiplet  
are charged under the $U(1)_R$ symmetry. 
For a Calabi-Yau GLSM,  the complexified FI parameters are the K\"ahler moduli of the Calabi-Yau while the complex parameters in the twisted superpotential correspond to the complex structure moduli.

\section{Localization of the Path Integral}
\label{sec:localization}

In this section we perform the exact computation of the partition function of the gauge theories constructed in the previous section. This requires   choosing a supercharge $\cQ$ in $SU(2|1)_A$ and a 
suitable deformation of the Lagrangian
\beq
\cL\rightarrow \cL +t \cQ V\,.
\eeq
By the familiar $t$-independence of the path integral (in favorable situations), the path integral reduces  to a one-loop  integral over the space of saddle points $\cM$ of $\cQ V$. The 
measure of integration is determined by classical action evaluated on the saddle points and by the   one-loop determinants $Z_{\text{1-loop}}$ of twisted vector and twisted chiral produced by the 
deformation term $\cQ V$. The contribution of the gauge fixing multiplet must also be included.

In formulas, for a collection of $\cQ$-invariant operators collectively denoted by $\cO$, we have that
\beq
\langle  \cO\rangle =\int_{\cM}  e^{-S|_\cM}\,   \cO|_{\cM}  \, Z_{\text{1-loop}}\,. 
\eeq
In this paper, $\cO$  is the two point function of a chiral operator $\cO_a$ at the north pole and an anti-chiral operator operator $\cO_{\bar a}$ at the south pole of the two-sphere.

\subsection{Choice of Supercharge and Decoupling Theorems}
\label{sec:deco}
We choose the following supercharge\footnote{We drop the index $A$, to avoid cluttering.} $\cQ$ in $SU(2|1)_A$ 
\begin{equation}
\label{cQ}
\cQ = S_{1} + Q_{2}\,.
\end{equation}
The $SU(1|1)$ subalgebra that $\cQ$ generates is
\begin{equation}
\label{su11}
  \cQ^{2}= J_{3} + \frac{R}{2} \hspace{50pt} \left[J_{3} + \frac{R}{2}, \cQ \right] =0 \,,
\end{equation}
where $J_3$ is a $U(1)$ isometry generator of $S^2$, and has two antipodal fixed points which we call the north and south poles of the two-sphere. 
$R$ is the $U(1)$ R-symmetry generator in $SU(2|1)_A$.

The (Grassmann even) Killing spinors \eqref{killl} parameterizing the transformations generated by $\cQ$ are
\begin{equation}
\label{cQKS}
\begin{aligned}
  \epsilon &= \exp\left(-\frac{i}{2} \theta \gamma^{\hat{2}} + \frac{i}{2} \varphi \right) \epsilon_{\circ}, \qquad \gamma^{\hat 3}\epsilon_{\circ}= \epsilon_{\circ}
  \\
  \bar{\epsilon}  &= \exp\left(+\frac{i}{2} \theta \gamma^{\hat{2}} - \frac{i}{2} \varphi \right) \bar{\epsilon}_{\circ}, \qquad \gamma^{\hat{1}}\bar{\epsilon}_{\circ} = \epsilon_{\circ}\,,
\end{aligned}
\end{equation}
where $(\theta,\varphi)$ are the canonical coordinates on $S^2$.

At the north pole of the two-sphere, gauge invariant operators $O_a(Y)$ constructed from   the lowest component of   twisted chiral multiplets are $\cQ$-invariant. Likewise, at the south pole, operators  
constructed from the lowest component of     twisted anti-chiral multiplets $O_{\bar{a}}(\bar{Y})$ are also $\cQ$-invariant. This follows from the supersymmetry transformation \rf{tcmsusy} generated 
by the spinors \rf{cQKS}. Therefore the two-point function
\beq
\langle \cO_a(Y)\, \cO_{\bar{b}}(\bar{Y})\rangle
\label{compuZamo}
\eeq
is $\cQ$-invariant and can be computed by supersymmetric localization.

  We now   prove that the two-sphere partition function  and two-point functions \rf{compuZamo} are independent of some of the parameters of the Lagrangian. First, we note that the twisted vector multiplet Lagrangian \rf{tvmL}
  as well as the FI and topological terms \rf{FItop} are all $\cQ$-exact. Explicitly
 \beq
 \label{QV1_tv}
   \cL_{\text{t.v.m.}}  = \frac{1}{4 g_{\text{YM}}^{2}} \cQ \widetilde{\cQ} \Tr\left( \eta \gamma^{\hat 3} \bar{\eta} + \frac{i}{r} \sigma \bar{\sigma} \right)- \nabla_{\mu} J^{\mu}_{\text{t.v.m.}}\,,
 \eeq
 where  $\widetilde{\cQ} = S_{1} - Q_{2}$, a supercharge in $SU(2|1)_A$ parametrized by Killing spinors \rf{cQKS} $-\epsilon$ and $\bar{\epsilon}$.\footnote{The total derivative terms are written down in appendix \ref{app:QV}.}
 Likewise
 \begin{equation}
\label{QV2}
\begin{aligned}
   \cL_{\text{FI}} &= \frac{\xi }{2 i } \cQ \Tr\left( \bar{\epsilon}\gamma^{\hat 3}\bar{\eta} + \epsilon \gamma^{\hat 3} \eta \right) - \nabla_{\mu} J^{\mu}_{\text{FI}}
  \\[10pt]
  \cL_{\text{top}} &= \frac{\vartheta}{4\pi} \cQ \Tr\left( \bar{\epsilon}\gamma^{\hat 3}\bar{\eta} - \epsilon \gamma^{\hat 3} \eta \right) - \nabla_{\mu} J^{\mu}_{\text{top}}\,.
\end{aligned}
\end{equation}
By virtue of  equation \rf{superFI}, this follows from the more general result that the superpotential $\cW$ couplings \rf{supergeneral} are $\cQ$-exact \cite{Doroud:2012xw,Benini:2012ui}.   
The twisted chiral Lagrangian \rf{tcmL} is also $\cQ$-exact
\begin{equation}
\begin{aligned}
\label{QV1_tc}
   \cL_{\text{t.c.m.}}  = \frac{1}{2} \cQ\widetilde{\cQ} \left( \bar{G}Y - \bar{Y}G + \frac{i}{r} \bar{Y} Y \right) - \nabla_{\mu} J^{\mu}_{\text{t.c.m.}}\,.
   \end{aligned}
\end{equation}
We note, however, that twisted superpotential couplings \rf{twistedW} are not $\cQ$-exact.

 This shows that the gauge theory two-sphere partition function and two-point functions \rf{compuZamo}   are independent of the gauge couplings $g_{\text{YM}}^{2}$ and of the complexified FI parameters $\tau$, but   
 depend on the complex parameters in the twisted superpotential. Gauge coupling independence implies that the two-sphere partition function of a gauge theory is a renormalization group invariant observable. 
 In particular, it coincides with the partition function of a SCFT theory in the  extreme infrared, where $g_{\text{YM}}^{2}\rightarrow \infty$. This is none other than the sought-after Calabi-Yau NLSM when the 
 gauge theory has a geometric phase. Moreover, the Zamolodchikov metric  \rf{metricB} of operators in the chiral ring of the $\cN=(2,2) $ SCFT can be exactly computed in the ultraviolet GLSM, as these    
 correlators  have images in the ultraviolet GLSM through \rf{compuZamo}.\footnote{In our choice of coordinates, where the infrared NLSM is described by twisted chiral multiplets, a chiral ring element in the 
 infrared SCFT is the lowest component of a twisted chiral superfield while   an operator in the conjugate ring is the lowest component of a twisted anti-chiral superfield.} In conclusion, a  gauge theory   on  
 the two-sphere computes  the 
K\"ahler potential and associated Zamolodchikov metric of the infrared SCFT. When the GLSM has a geometric phase, the gauge theory computes these quantities 
for the complex structure moduli space of the Calabi-Yau.

\subsection{\texorpdfstring{$\cQ$}{Q}-Exact Deformation Term}

We proceed by deforming the gauge theory action by a $\cQ$-exact term
\beq
\cL \rightarrow \cL+t \cQ \cV\,.
\eeq
Following our discussion in the previous subsection, we can take 
\begin{equation}
\label{V}
  \cV =   \frac{1}{4 g^{2}} \widetilde{\cQ} \Tr\left( \bar{\eta}\gamma^{\hat 3}\eta + \frac{i}{r} \sigma\bar{\sigma} \right) 
	  -\frac{i}{2} \chi \Tr(\bar{\epsilon}\gamma^{\hat 3}\bar{\eta} + \epsilon \gamma^{\hat 3} \eta) + \frac{1}{2} \widetilde{\cQ} \left( \bar{G} Y  - \bar{Y} G + \frac{i}{r} \bar{Y} Y  
	  \right)  \,.
\end{equation}
The bosonic part of $\cQ\cV$ can be recast into the positive definite form
\begin{equation}
\label{QVbosonic}
  \frac{1}{2 g^{2}} \Tr\left( |D_{\mu}\sigma|^{2} + \frac{1}{4}[\sigma,\bar{\sigma}]^{2} + F^{2} + \widetilde{\rmD}^{2} \right)+ \left| D_{\mu}Y \right|^{2} 
	+ \left| G \right|^{2} + \frac{1}{2}\left( |\sigma Y|^{2} + |\bar{\sigma}Y|^{2}\right) + \frac{g^{2}}{2} \left(Y\bar{Y} - \chi \right)^{2}\,,
\end{equation}
where $\widetilde{\rmD} = \rmD + i g^{2} (Y\bar{Y} - \chi)$.   Positive definiteness follows from the reality conditions
\begin{equation}
\label{rcs}
\begin{aligned}
  \sigma^{\dagger} &= \bar{\sigma}
  \\[3pt]
  Y^{\dagger} &= \bar{Y}
\end{aligned}
\hspace{50pt}
\begin{aligned}
  \widetilde{\rmD}^{\dagger} &= \widetilde{\rmD}
  \\[3pt]
  G^{\dagger} &= \bar{G}
\end{aligned}
\hspace{50pt}
\begin{aligned}
  F^{\dagger} &= F
  \\
  A_{\mu}^{\dagger} &= A_{\mu}\,.
\end{aligned}
\end{equation}

By adding this deformation term to the action  and taking the limit $t\rightarrow\infty$, we are able to apply the saddle point method, which is exact,  and localize the path integral 
to the extrema of $\cQ\cV$. Since the bosonic part of the deformation term is positive definite, all the paths that contribute to the  path integral  lie at the global minimum surface $\cQ\cV=0$ in the space of fields.
The space of saddle points that we must integrate over in the path integral is therefore\footnote{For $\chi=0$, $\sigma$ can be non-zero, but then at least one $Y$ must vanish. The fermionic superpartner of this field, 
however, has a fermionic  zero mode, and this saddle point does not contribute.}
\begin{equation}
\label{higgsbranch}
\cM=	\left\{Y | Y=Y_{\circ},\ Y_{\circ}\bar{Y}_{\circ} - \chi =0\right\}/G_{\text{global}}\,,
\end{equation}
with  all the other fields vanishing. $Y=Y_{\circ}$ is constant on the two-sphere.\footnote{Even though the parameter $\chi$ enters in the definition of $\cM$, we shall prove  that the partition function is 
independent of $\chi$,  as it should, since it is  the coefficient of a $\cQ$-exact term in \rf{V}.} 
 Field configurations related by the residual gauge transformation $G_{\text{global}}$ (the global part of the gauge group $G$)  must be identified
\beq
Y_{\circ}\simeq e^{i\alpha} Y_{\circ}\,,
\eeq
where $\alpha$ acts on $Y$ in the corresponding representation ${\bf R}$ of the gauge group. $\cM$ is therefore the K\"{a}hler quotient space
 \beq
 \cM= \mathbb{C}^{|\bf R|}//G_{\text{global}}\,.
 \label{quotient}
 \eeq 

Localization has to be performed for the gauge fixed functional integral (see appendix \ref{app:BRST} for details). For the background field configurations \eqref{higgsbranch}, we fix the Lorenz gauge which is
compatible with $A_{\mu}=0$. For the field fluctuations in the computation of the one-loop determinant however, it is much more convenient to fix an $R_\xi$-like  gauge adapted to the Higgs phase of the theory.  
This requires introducing  gauge fixing terms and a fermionic generator $\cQ_{\text{BRST}}$. We localize the path integral with respect to the BRST deformed supercharge $\hat{\cQ} = (\cQ+\cQ_{\text{BRST}})$ using
  as the deformation term $\hat{\cQ}\cV^{\prime}$, where $\cV^{\prime} = \cV + \cV_{\text{G.F.}}$. The space of saddle points of the gauge fixed theory remains unaffected by the inclusion of the gauge fixing terms, however,
the gauge fixing terms play an important role in the computation of the measure factor $Z_{\text{1-loop}}$.

\subsection{Partition Function and Zamolodchikov Metric}
\label{sec:measure}

Calculation of the measure of integration in the space of saddle points $\cM$ requires computing the one-loop determinant $Z_{\text{1-loop}}$ of twisted vector, twisted chiral and ghost multiplets  
around the saddle point configurations $\cM$. This is achieved by integrating out to quadratic order in the fluctuations the   deformation and gauge fixing terms $\hat{\cQ}\cV^{\prime}$.

Consider a gauge theory with gauge group $G=U(1)^{N_{c}}$ coupled to $N_{f}$ twisted chiral multiplets 
with charges $Q^{a}_{I}$ under $U(1)^{N_{c}}$, where $a=1,\dots,N_{c}$ and $I=1,\dots,N_{f}$. Supersymmetry on the two-sphere requires   anomaly cancelation for the $U(1)_R$ R-symmetry, which yields the constraints 
\beq
\sum_I Q^a_I=0 \qquad a=1,\dots,N_{c}\,.
\label{anoma}
\eeq
The one-loop determinant around the saddle points \rf{higgsbranch} is given by the determinant of an $N_c\times N_c$ matrix (see appendix \ref{app:1-loop} for details)
\begin{equation}
Z_{\text{1-loop}} = \det(M^{\dagger}M)\,.
\end{equation}
Here $M$ is an  the $N_{f}\times N_{c}$ mass matrix and $M^{\dagger}$  is its  hermitian conjugate. They are given by\footnote{We drop the subindex of $Y_{\circ}$ in order to avoid cluttering.}
\begin{equation}
\label{Ms}
  M^{\ a}_{I} = Q_{I}^{a}Y_{I}\,,\qquad M^{\dagger\, I}_{a} = Q^{a}_{I}\bar{Y}_{I}\,.
\end{equation}
We note that $N_f\ge N_c$ is a necessary condition for the matrix $M^{\dagger}M$ to be non-degenerate. For $N_{f}<N_{c}$, there is a linear combination of the $U(1)$ generators under which all the 
twisted chiral fields are neutral, and the associated gaugino has a fermionic zero mode,\footnote{Given by $\lambda=\bar\epsilon$, where $\bar\epsilon$ is the conformal Killing spinor \rf{cQKS}.} 
and therefore the path integral vanishes.
 
 Evaluating the classical action and operator insertions on the saddle points we obtain\footnote{As explained in \cite{Doroud:2012xw},
 the partition function is also proportional to   $r^{c/3}$, due to the usual conformal anomaly,  where $c$ is the central charge.}
 \beq
 \langle \cO_a(N) \cO_{\bar b}(S)\rangle =\int \text{vol}_{\cM}\cO_a(Y) \cO_{\bar b}(\bar Y) \,Z_{\text{1-loop}} \,e^{r W(Y)-r\overline{W}(\bar Y)}\,,
 \label{partii}
 \eeq
where $\text{vol}_{\cM}$ is the volume form on the space of saddle points $\cM$ \rf{higgsbranch}.
The volume form on $\cM$, which is the    quotient space \rf{quotient}, can be written in terms of the volume form of the ambient flat space $\mathbb{C}^{N_{f}}$ by inserting
appropriately normalized Dirac delta distributions and dividing by the volume of the $U(1)^{N_c}$ gauge orbits:
\begin{equation}
\label{voal}
  \text{vol}_\cM = {\rmd^{N_{f}} Y \wedge \rmd^{N_{f}} \bar{Y}\over \text{vol}(G_{\text{global}})} \,  \det\left[\frac{\partial F_a}{\partial Y_b}\right] \prod_{a}\delta\left(F_a \right)\,,
\end{equation}
where
\beq
F_a=\sum_{I}Q^{a}_{I} |Y^{I}|^{2}-\chi_a\,.
\eeq

On the ambient space $\mathbb{C}^{N_{f}}$, we can define the Hamiltonian action of the complexification $U(1)^{N_c}_\mathbb{C}$ of the gauge group. The vector fields that generate the real gauge transformations are
\beq
\label{rho}
  \rho_{a} = i\sum_{I} Q^a_{I} \left( Y_{I}\partial_{I} - \bar{Y}_{I} \bar{\partial}_{I} \right)\qquad a=1,\ldots,N_c\,,
\end{equation}
while 
\beq
\label{mu}
v_a=-\sum_{I} Q^a_{I} \left( Y_{I}\partial_{I} + \bar{Y}_{I} \bar{\partial}_{I} \right)\qquad a=1,\ldots,N_c\,,
\eeq
generate imaginary gauge transformations. They act, respectively, as
\beq
Y_I\rightarrow e^{i\sum_{a}Q^{a}_{I} \tau_{1}^{a}} Y_I\,,\qquad Y_I\rightarrow e^{-\sum_{a} Q^{a}_{I}\tau_{2}^{a}} Y_I\qquad \tau_{1},\tau_{2} \in \mathbb{R}\,.
\eeq
The  moment map associated with the  imaginary transformation generated by the  $a$-th $U(1)$ factor in the gauge group is given by 
\beq
\label{momentt}
\mu_{a} = -\frac{1}{2}\sum_{I}Q^{a}_{I} |Y^{I}|^{2}\qquad a=1,\ldots,N_c\,,
\eeq
as it obeys
\beq
\rmd \mu_a= \imath_{v_a}\omega\,,
\eeq
where $\omega$ is the K\"ahler form in $\mathbb{C}^{N_{f}}$. Therefore, the $\rmD$-term equations entering in the definition of $\cM$ in \rf{higgsbranch} 
\beq
\left\{ \sum_{I}Q^{a}_{I} |Y^{I}|^{2}=\chi  \longleftrightarrow F_a=0\,;\, a=1,\ldots,N_c \right\}\,,
\eeq
can be interpreted as the moments maps for the imaginary gauge transformations
\beq
2\mu_a+\chi=0\qquad a=1,\ldots,N_c\,.
\eeq
We note that these moment maps  obey the   equations
\beq
 \rmd\mu_{a}\cdot \rmd\mu_{b} =  \left(M^{\dagger}M\right)_{ab}\,,
 \label{cosa}
\eeq
where $\rmd$ denotes the exterior derivative and the inner product $\rmd\mu_{a}\cdot \rmd\mu_{b}$ is the $\mathbb{C}^{N_{f}}$ inner product. As a direct consequence of the anomaly cancellation conditions \rf{anoma},
the holomorphic and anti-holomorphic factors in the the measure
\beq
\rmd^{N_{f}} Y \wedge \rmd^{N_{f}} \bar{Y}
\eeq
are each invariant under the complexified gauge transformations $U(1)^{N_c}_\mathbb{C}$. Furthermore, the twisted superpotential $W(Y)$ and $\overline{W}(\bar Y)$ are also invariant under complex gauge transformations, whereas 
$Z_{\text{1-loop}}$ is only invariant under real gauge transformations. This observation suggests a change of coordinates $\{Y\}\rightarrow\{X,\tau\}$, to some gauge invariant coordinates $X$ and the (complex) gauge orbit
coordinates $\tau$, where the integration over the complex gauge orbits is localized to the real gauge orbits due to the $\delta$-distributions arising from the $\rmD$-term equations.

In computing the volume form $\text{vol}_{\cM}$ we must quotient by the volume of the orbit of $U(1)^{N_c}$   real gauge transformations. It follows from \rf{rho} that it is given by
\beq
\text{vol}(G_{\text{global}})=(2\pi)^{N_c} \det\left(\rho_{a}\cdot\rho_{b}\right)^{1/2} \,.
\eeq
By virtue of \rf{rho} we have that
\beq
\rho_{a} \cdot \rho_{b} = 4\left(M^{\dagger}M\right)_{ab}\,,
\eeq
which combined with \rf{cosa} implies\footnote{The Jacobian factor $J_{\{b\}}=\det\left(\partial F_{a}/\partial Y_{b}\right)$ in \eqref{voal} assumes that one carries out the integration over the $Y_{\{b\}}$ planes first, 
treating $Y_{I}$ as constant for $I\neq b$. More covariantly, one may write $J=\sqrt{\det\left( \rmd F_{a} \cdot \rmd F_{b} \right)}$ which takes the order of integration into account.}
that the Jacobian appearing with the delta functions in \rf{voal} precisely cancels with the volume of the gauge orbit.
   
Altogether,  the correlator \rf{partii}   can be written as 
\begin{equation}
\label{correlator}
  \langle \cO_a(N) \cO_{\bar b}(S)\rangle = \int {\rmd^{N_{f}} Y \wedge \rmd^{N_{f}} \bar{Y}\over (2\pi)^{N_{c}}} \cO_a(Y)\, \cO_{\bar b}(\bar Y)\, 
  \det\left(M^{\dagger}M\right) \prod_{a}\delta\left(2\mu_{a}+\chi_{a} \right)\,  e^{rW(Y)-r\overline{W}(\bar Y)}\,,
\end{equation}
with $M$ and $M^\dagger$ defined in \rf{Ms} and $\mu_a$ in \rf{momentt}. The partition function is obtained by placing the identity operator at the north and south poles of the two-sphere, yielding
\begin{equation}
\label{PF}
  \cZ_{B} = \int {\rmd^{N_{f}} Y \wedge \rmd^{N_{f}} \bar{Y}\over (2\pi)^{N_{c}}}\, 
  \det\left(M^{\dagger}M\right) \prod_{a}\delta\left(2\mu_{a}+\chi_{a} \right)\,  e^{rW(Y)-r\overline{W}(\bar Y)}\,.
\end{equation}

\section{Calabi-Yau Geometries}
\label{sec:CYG}

The two-sphere partition function \eqref{PF} of a Calabi-Yau GLSM is expected to compute  the K\"{a}hler potential $\cK_{C}$ for the complex structure moduli of the 
corresponding Calabi-Yau manifold. Concretely, we expect
\begin{equation}
  \label{ZB=KC}
  \cZ_{B} = e^{-\cK_{C}} = i^{\,\text{dim}M} \int_M \Omega\wedge\overline{\Omega}\,,
\end{equation}
where $\Omega$ is the nowhere vanishing holomorphic top form of the corresponding Calabi-Yau. We now turn to explicitly demonstrating this for various families of Calabi-Yau geometries.

\subsection{Quintic Hypersurfaces in \texorpdfstring{$\mathbb{CP}^{4}_{[Q_1,\ldots,Q_5]}$}{CP4[Q]}}

Consider the partition function \eqref{PF} in the case of a $U(1)$ gauge theory coupled to five twisted chiral multiplets $Y_{I}$ with charges $Q_{I}$ and a twisted chiral multiplet $P$ with charge $-q$.
The anomaly cancellation condition requires the sum of the charges of all of the twisted chiral multiplets vanish, \emph{i.e.}
\begin{equation}
\label{acc}
  q = \sum_{I}Q_{I}\,.
\end{equation}
The twisted superpotential for GLSMs corresponding quintic hypersurfaces in $\mathbb{CP}^{4}_{[Q_1,\ldots,Q_5]}$ has the general form
\begin{equation}
\label{quinticsuperpotential}
  W = P G_{5}(Y) \,,
\end{equation}
where $G_{5}(Y)$ is a transverse polynomial satisfying  
\begin{equation}
  G_{5}( \lambda^{Q_{I}} Y_{I} ) = \lambda^{q} G_{5}(Y)\qquad  \lambda\in \mathbb{C}^*\,.
\end{equation}
The two-sphere partition function takes the form\footnote{We set $r=1$ from now on.}
\begin{equation}
\label{PF_quintic1}
  \cZ = \frac{1}{2\pi}\int \rmd^{5} Y \wedge \rmd^{5} \bar{Y}\wedge \rmd P\wedge\rmd\bar{P} \ M^{\dagger}M\ \delta\left(2\mu+\chi\right) e^{W-\overline{W}}\,,
\end{equation}
  where the moment map and the mass matrix are given by
\begin{equation}
\label{mmquintic1}
\begin{aligned}
  -2\mu &= \sum_{I} Q_{I}|Y_{I}|^{2} - q |P|^{2}\,,
  \\[5pt]
  M^{\dagger}M &= \sum_{I} Q_{I}^{2}|Y_{I}|^{2} + q^{2} |P|^{2}\,.
\end{aligned}
\end{equation}
We remark the the anomaly cancellation condition \eqref{acc} guarantees that the flat measure and the twisted superpotential factor in \eqref{PF_quintic1} are invariant under global \emph{complex} gauge transformation.
It is therefore natural to consider the change of variables
\begin{equation}
\label{cov1_quintic}
\begin{aligned}
  Y_{I} &= e^{iQ_{I}\tau}x_{I}\,,
  \\[3pt]
  P &= e^{-iq\tau}p\,,
\end{aligned}
\end{equation}
with $x_{5}=\text{constant}$. In these coordinates, complex gauge transformations act only as a shift of the $\tau$ coordinate and therefore $\tau$ is the (complex) gauge orbit coordinate.
The invariance of the ambient space volume form and the twisted superpotential under complex gauge transformations generated by $\partial_{\tau}$ becomes manifest in the new coordinates. 
The volume form of $\mathbb{C}^{6}$ in the new coordinates is
\begin{equation}
  \rmd^{5} Y \wedge \rmd^{5} \bar{Y}\wedge \rmd P\wedge\rmd\bar{P} = Q_{5}^{2}\,|x_{5}|^{2}\ \rmd^{4} x \wedge \rmd^{4} \bar{x}\wedge \rmd p \wedge\rmd \bar{p} \wedge \rmd \tau\wedge\rmd\bar{\tau}\,.
\end{equation}
while the twisted superpotential retains it's original form
\begin{equation}
  W = P G_{5}(Y) = p\, G_{5}(x)\,.
\end{equation}
The moment map and the mass matrix \eqref{mmquintic1}, however, depend explicitly of the imaginary $\tau$ direction, denoted by  $\tau_{2}$, as they are only invariant under real gauge transformations,
and may be rewritten as
\begin{equation}
\label{mmquintic2}
\begin{aligned}
  -2\mu &= \sum_{I=1}^{5} Q_{I}\, e^{-2Q_{I}\tau_{2}} |x_{I}|^{2} - q\, e^{2 q \tau_{2}}|p|^{2}\,,
  \\[5pt]
  M^{\dagger}M &= \sum_{I=1}^{5} Q_{I}^{2}\,e^{-2Q_{I}\tau_{2}} |x_{I}|^{2} + q^{2} e^{2 q \tau_{2}} |p|^{2}\,.
\end{aligned}
\end{equation}
The partition function \eqref{PF_quintic1} in the new coordinates is
\begin{equation}
\label{PF_quintic2}
  \cZ = -2i Q_{5}^{2} \, |x_{5}|^{2} \int \rmd^{4} x \wedge \rmd^{4} \bar{x}\wedge \rmd p \wedge \rmd\bar{p}\left( e^{p\,G_{5}(x)-\bar{p}\, \bar{G}_{5}(\bar{x})} 
  \int \rmd\tau_{2} \ M^{\dagger}M\ \delta\left(2\mu+\chi\right) \right) \,,
\end{equation}
where we have carried out the integration over $\tau_{1}$ which only contributes a factor of $2\pi$. It is clear from \eqref{mmquintic2} that $M^{\dagger}M$ is the Jacobian $\partial_{\tau_{2}}\mu$, that is
\begin{equation}
  \rmd\tau_{2} M^{\dagger}M = \rmd\tau_{2} \frac{\partial \mu}{\partial\tau_{2}}\ \dot{=}\ \rmd\mu
\end{equation}
keeping $x_{I}$ and $p$ constant. This implies that the integration over $\tau_{2}$ can be readily carried out yielding\footnote{In general, the equations $2\mu+\chi=0$ have multiple solutions for $\tau_{2}$; this only introduces
a multiplicative factor which we ignore.}
\begin{equation}
  \int\rmd\tau_{2}\, M^{\dagger}M \delta(2\mu+\chi)\ \dot{=}\ \int \rmd\mu\, \delta(2\mu+\chi) =1/2\,.
\end{equation}
The partition function \eqref{PF_quintic2} can be put into the proposed form in terms of an integral over the holomorphic three-form by performing the integration over the complex variables  $p$ and $\bar{p}$ as well as
the integration over one of the $x$ planes, say $x_{4}$. Integrating over $p$ imposes the embedding equation $G_{5}=0$ in $\mathbb{CP}^{4}_{[Q_1,\ldots,Q_5]}$  via $\delta$ distributions 
\begin{equation}
\label{deltaG5}
 \int \rmd p \wedge\rmd\bar{p}\ e^{p\,G_{5} - \bar{p}\,\bar{G}_{5}} = - \frac{1}{4\pi^{2}} \delta(G_{5}) \delta(\bar{G}_{5})
\end{equation}
and finally, integrating over $x_{4}$ and $\bar{x}_{4}$ yields
\begin{equation}
\label{PF_quintic}
  \cZ =  i\frac{ Q_{5}^{2}\,|x_{5}|^{2}}{4\pi^{2}} \sum_{\substack{\{x_{4}|G_{5}=0\} \\ \{\bar{x}_{4}|\bar{G}_{5}=0\}}}\int \frac{\rmd x_{1} \wedge \rmd x_{2} \wedge \rmd x_{3}}{\partial_{4}G_{5}(x)}
  \wedge \frac{\rmd \bar{x}_{1} \wedge \rmd \bar{x}_{2} \wedge \rmd \bar{x}_{3}}{\bar{\partial}_{4}\bar{G}_{5}(\bar{x})} \,.
\end{equation}
From \eqref{PF_quintic} we can read off the holomorphic three-form to be
\begin{equation}
\label{quintic_Omega}
  \Omega = \frac{Q_{5}}{2\pi}\ \frac{x_{5}\,\rmd x_{1} \wedge \rmd x_{2} \wedge \rmd x_{3}}{\partial_{4}G_{5}(x)}\,,
\end{equation}
which matches the well known formulae for the holomorphic three-form presented in \cite{Strominger:1985it,Candelas:1987kf} of quintic hypersurfaces in $\mathbb{CP}^{4}_{[Q_1,\ldots,Q_5]}$.
We remark that although \eqref{quintic_Omega} appears to have singularities whenever $\partial_{4}G_{5}=0$, via a simple change of coordinates, corresponding to integrating \eqref{deltaG5} 
with respect to $x_{1}$ instead of $x_{4}$, it may be written as
\begin{equation}
  \Omega = - \frac{Q_{5}}{2\pi}\ \frac{x_{5}\,\rmd x_{2} \wedge \rmd x_{3} \wedge \rmd x_{4}}{\partial_{1}G_{5}(x)}\,.
\end{equation}
Since the polynomial $G_{5}(x)$ is transversal and $x_{5}\neq0$, it follows that the holomorphic three-form $\Omega$ is non-singular and nowhere vanishing.

\vskip+8pt

\noindent{\it Mirror Quintic Complex Structure K\"ahler Potential}
\vskip+8pt

In \cite{Gomis:2012wy} the $SU(2|1)_A$-invariant partition function for the familiar quintic three-fold in $\mathbb{CP}^4$ was shown to coincide with the  $SU(2|1)_A$-invariant  partition function of the Hori and Vafa mirror theory \cite{Hori:2000kt}. 
This is 
a $U(1)$ vector multiplet coupled to twisted chiral multiplets $(Y_1,\ldots,Y_5,Y_P)$   with a 
twisted superpotential 
\begin{align}
  W =    \bigg[i \, \Sigma \left( \sum_{a=1}^5 Y^a - 5 Y_P + 2\pi i \tau \right)
  - \Big( \sum_{a=1}^5 e^{-Y^a} + e^{-Y_P} \Big) \bigg]\,,
\end{align}
where $\Sigma$ is the field strength multiplet. As shown in \cite{Gomis:2012wy}, the relation to the Mellin-Barnes like formula for $SU(2|1)_A$ invariant gauge theories derived in \cite{Doroud:2012xw,Benini:2012ui} follows by integrating out the twisted chiral multiplet fields. Explicitly, decomposing the integral into contours\footnote{$C$ is the Hankel contour, which starts at  $-\infty-i\epsilon$, then goes around the branch cut along the negative real $t$
axis, and ends up at $\infty+i\epsilon$.}
\beq
\int_{Y^*=\bar Y} dY d\bar Y e^{-e^{-Y}+iQ\Sigma Y} e^{e^{-\bar Y}+i Q \bar \Sigma \bar Y}=\int_{0}^\infty dt\, e^{-t}\, t^{-iQ\Sigma-1}\int_C dt\, e^{\bar t}\, {\bar t}^{-iQ\bar \Sigma-1}\,,
\eeq
and with the identities
\beq
\int_{0}^\infty dt\, e^{-t}\, t^{-iQ\Sigma-1}=\Gamma(-iQ\Sigma)\,,  \qquad \ \ \ \int_C dt\, e^{\bar t}\, {\bar t}^{-iQ\bar \Sigma-1}={2\pi i \over \Gamma(1+iQ\bar \Sigma)}\,,
\eeq
 one arrives at the gauge theory result \cite{Gomis:2012wy}.\footnote{This is a streamlined version of the identity derived in \cite{Gomis:2012wy}.}

The two-sphere partition function of the mirror theory can be reduced to   an orbifold Landau-Ginzburg model by integrating out $\Sigma$. This yields \cite{Gomis:2012wy}
\begin{align}
  \cZ_{\text L.G.} = \int \prod_{a=1}^5 d {\widetilde X}_a d \overline{{\widetilde X}}_a   \ e^{- W_{\text{eff}}+ \overline{W}_{\text{eff}}} \,,
\end{align}
where the effective twisted superpotential is
\begin{align}
W_{\text{eff}} = \sum_a {\widetilde X}_a^5 + e^{-2\pi i \tau/5}\prod_a {\widetilde X}_a\ .
\label{quinticW}
\end{align}
The canonical  variables ${\widetilde X}_a$ are given by
\begin{align}
  {\widetilde X}_a = e^{-\frac15 Y_a} \,,
\end{align}
and therefore we must orbifold by
\begin{align}
\widetilde X_a \simeq e^{2\pi i /5} \widetilde X_a\,.
\end{align}
This orbifold Landau-Ginzburg model  realizes the mirror Calabi-Yau geometry: the mirror quintic $W$. Indeed, it is easy to show that the  orbifold Landau-Ginzburg model  partition function also computes the K\"ahler potential on the complex structure moduli space of the mirror quintic $W$
\beq
 \cZ_{\text L.G.}=i \int_W \Omega\wedge \overline \Omega\,.
\eeq

\subsection{Complete Intersection Surfaces in \texorpdfstring{$\mathbb{CP}^{n}_{[Q_{1},\dots,Q_{n+1}]}$}{CP(n)}}

The analysis of the last section can be easily generalized to intersection of multiple hypersurfaces in $\mathbb{CP}^{n}_{[Q_{1},\dots,Q_{n+1}]}$. 
As the analysis is quite parallel to that of the last section, some details are omitted here.

Consider the partition function \eqref{PF} in the case of a $U(1)$ gauge theory, this time coupled to $n+1$ twisted chiral multiplets $Y_{I}$ with charges $Q_{I}$ and $m$ twisted chiral multiplet $P_{\alpha}$ 
with charges $-q_{\alpha}$. Imposing the anomaly cancellation condition restricts the charges to satisfy
\begin{equation}
  \sum_{\alpha} q_{\alpha} = \sum_{I}Q_{I}\,.
\end{equation}
The partition function takes the form
\begin{equation}
\label{PF_n-m1}
  \cZ = \frac{i^{n+m+1}}{2\pi}\int \rmd^{n+1} Y \wedge \rmd^{n+1} \bar{Y}\wedge \rmd^{m} P\wedge\rmd^{m}\bar{P} \ M^{\dagger}M\ \delta\left(2\mu+\chi\right) e^{W-\overline{W}}
\end{equation}
where the twisted superpotential is linear in $P_{\alpha}$ and is a polynomial in $Y_{I}$,
\begin{equation}
\label{n-msuperpotential}
  W= \sum_{\alpha}P_{\alpha}G_{\alpha}(Y)\,,
\end{equation}
with the polynomials $G_{\alpha}$ satisfying
\begin{equation}
  G_{\alpha}(\lambda^{Q_{I}}Y_{I}) = \lambda^{q_{\alpha}}G_{\alpha}(Y)\,.
\end{equation}
We emphasize again that both the twisted superpotential term and the volume form for the ambient space $\mathbb{C}^{n+1+m}$ are invariant under complex gauge transformations. The change of variables
\begin{equation}
\begin{aligned}
  Y_{I} &= e^{iQ_{I}\tau} x_{I}\,,
  \\
  P_{\alpha} &= e^{-iq_{\alpha}\tau} p_{\alpha}\,,
\end{aligned}
\end{equation}
with $x_{n+1}=\text{constant}$, makes this invariance manifest as the gauge transformations in the new coordinates act simply as a shift in $\tau$. The twisted superpotential in the new coordinates assumes the $\tau$-independent form
\begin{equation}
  W= \sum_{\alpha} p_{\alpha} G_{\alpha}(x)
\end{equation}
and the volume form is
\begin{equation}
  2i^{n+m}Q_{n+1}^{2}\,|x_{n+1}|^{2} \rmd^{n}x \wedge\rmd^{n}\bar{x} \wedge \rmd^{m}p\wedge\rmd^{m}\bar{p} \wedge \rmd\tau_{1}\wedge\rmd\tau_{2}\,.
\end{equation}
Here $\tau_{1}$ and $\tau_{2}$ are the real and imaginary parts of the $\tau$ coordinate parameterizing the compact and non-compact directions of the gauge orbit surface.
The moment map, which has an explicit $\tau_{2}$ dependence takes the form
\begin{equation}
\label{mmn-m2}
  -2\mu = \sum_{I} e^{-2Q_{I}\tau_{2}} Q_{I}|x_{I}|^{2} - \sum_{\alpha} q_{\alpha} e^{2q_{\alpha}\tau_{2}}p_{\alpha}\,,
\end{equation}
while $M^{\dagger}M$ can be related to the moment map, as in the case of quintic hypersurfaces, by $\tau_{2}$ differentiation of the latter,
\begin{equation}
  M^{\dagger}M = \frac{\partial \mu}{\partial\tau_{2}}\,.
\end{equation}
The integration over $\tau_{1}$ and $\tau_{2}$ may then be carried out as was done for quintic hypersurfaces \eqref{PF_quintic2}, yielding
\begin{equation}
\label{PF_n-m2}
\begin{aligned}
  \cZ &= \frac{i^{n+m}}{\pi}Q_{n+1}^{2}\,|x_{n+1}|^{2} \int \rmd^{n} x \wedge \rmd^{n} \bar{x}\wedge \rmd^{m}p \wedge \rmd^{m}\bar{p}\left( e^{\sum_{\alpha}(p_{\alpha}G_{\alpha}-\bar{p}_{\alpha}\bar{G}_{\alpha})}
  \int\rmd\tau_{1}\wedge\rmd\tau_{2} \frac{\partial \mu}{\partial\tau_{2}} \ \delta\left(2\mu+\chi\right) \right)
  \\
  &=i^{n+m}Q_{n+1}^{2}\,|x_{n+1}|^{2} \int \rmd^{n} x \wedge \rmd^{n} \bar{x}\wedge \rmd^{m}p \wedge \rmd^{m}\bar{p}\ e^{\sum_{\alpha}(p_{\alpha}G_{\alpha}-\bar{p}_{\alpha}\bar{G}_{\alpha})}\,.
\end{aligned}
\end{equation}
This is a simple generalization of the case of a hypersurface defined by a single embedding equation studied in the last section, with multiple $p$ fields, one for each constraint. Integration over the $p$ planes then
imposes all the constraints leading to
\begin{equation}
\label{PF_n-m3}
  \cZ = \frac{i^{n-m}Q_{n+1}^{2}\,|x_{n+1}|^{2}}{(2\pi)^{2m}} \int \rmd^{n} x \wedge \rmd^{n} \bar{x}\prod_{\alpha} \delta(G_{\alpha})\delta(\bar{G}_{\alpha})\,.
\end{equation}
Carrying out the integration over the $m$ dimensional space $\{x_{I}|I=n-m+1,\dots,n\}$ we arrive at the desired expression
\begin{equation}
\label{PF_n-m}
  \cZ = i^{n-m} \frac{Q^{2}_{n+1}\,|x_{n+1}|^{2}}{(2\pi)^{2m}} \sum_{ \substack{ \{x_{n-m+\beta}|G_{\alpha}=0\} \\ \{\bar{x}_{n-m+\beta}|\bar{G}_{\alpha}=0\} } } 
  \int \frac{ \rmd x_{1}\wedge \dots \wedge \rmd x_{n-m}}{\det\left(\partial_{n-m+\beta} G_{\alpha}(x)\right)} \wedge 
  \frac{ \rmd \bar{x}_{1}\wedge \dots \wedge \rmd \bar{x}_{n-m}}{\det\left(\bar{\partial}_{n-m+\beta} \bar{G}_{\alpha}(\bar{x})\right)}\,,
\end{equation}
where each determinant in the denominator is computed over the $\alpha$ and $\beta$ indices. This yields the holomorphic $n-m$ form
\begin{equation}
\label{Omega_n-m}
  \Omega= \frac{Q_{n+1}}{(2\pi)^{m}}\ \frac{x_{n+1}\, \rmd x_{1}\wedge \dots \wedge \rmd x_{n-m}}{\det\left(\partial_{n-m+\beta} G_{\alpha}(x)\right)}\,
\end{equation}
for the intersection of $m$ hypersurfaces in $\mathbb{CP}^{n}_{[Q_{1},\dots,Q_{n+1}]}$ \cite{Strominger:1985it,Candelas:1987kf}. That $\Omega$ appears to be singular whenever $\det\left(\partial_{n-m+\beta} G_{\alpha}(x)\right)=0$ is an artifact of the choice of
coordinates. For these points on the manifold, there is a different choice $\{x_{\sigma(\alpha)},\alpha=1,\dots,m\}$ of coordinates to integrate the $\delta$-distributions in \eqref{PF_n-m3}, 
such that \eqref{Omega_n-m} is non-singular.

\subsection{Complete Intersection of Hypersurfaces in Product of Weighted Projective Spaces}
\label{CIHPWPS}

As a much more general class of complete intersections with abelian GLSM realization, we now consider consider the partition function \eqref{PF} in the case of $U(1)^{N_{c}}$ gauge theory
with $N_{f}=n+m+N_{c}$ twisted chiral multiplets $Y_{I}$ with charge matrix $\{Q^{a}_{I}|a=1,\dots,N_{c};\,I=1,\dots,N_{f}\}$. The anomaly cancellation conditions restricts the charge matrix to obey
\begin{equation}
  \sum_{I} Q^{a}_{I} = 0\quad\text{for all}\ a\,.
\end{equation}
The partition function has the general form \eqref{PF}
where the superpotential is a polynomial in $\{X_{I} = Y_{I}| I = 1,\dots, N_{f}-m\}$  and is linear in $\{P_{\alpha} = Y_{\alpha}|\alpha=N_{f}-m+1,\dots,N_{f}\}$,
\begin{equation}
  W= \sum_{\alpha}P_{\alpha}G_{\alpha}(X)\,.
\end{equation}
The polynomials $G_{\alpha}$ satisfy
\begin{equation}
  G_{\alpha}(\lambda^{Q^{a}_{I}}X_{I}) = \lambda^{-Q^{a}_{\alpha}}G_{\alpha}(X)\,,
\end{equation}
which guarantees the invariance of the twisted superpotential under $U(1)^{N_{c}}_{\mathbb{C}}$ gauge transformations. 
As before, we introduce the complex $\tau^{a}$ coordinates, one for each $U(1)$ factor in the gauge group, via
\begin{equation}
\begin{aligned}
  X_{I} &= e^{i\sum_{a}Q^{a}_{I}\tau^{a}} x_{I}\,,
  \\
  P_{\alpha} &= e^{i\sum_{a}Q^{a}_{\alpha}\tau^{a}} p_{\alpha}\,,
\end{aligned}
\end{equation}
and with\footnote{This amounts to choosing inhomogeneous coordinates on the Calabi-Yau.} $x_{n+1}=\dots=x_{n+N_{c}}=1$. This isolates the action of each $U(1)_{a}$ factor in the gauge group to a shift in 
$\tau_{a}$ and highlights the gauge invariance of the twisted superpotential
\begin{equation}
  W= \sum_{\alpha} p_{\alpha} G_{\alpha}(x)\,.
\end{equation}
To write the volume form of $\mathbb{C}^{N_{f}}$ in the new coordinates, first consider the volume form of the subspace $\mathbb{C}^{N_{c}}$ of constant $x_{I}$. The holomorphic part of this volume form
may be written as
\begin{equation}
  \rmd X_{n+1}\wedge\dots\wedge\rmd X_{n+N_{c}} = i^{N_{c}} \sum_{a_{1},\dots,a_{N_{c}}} Q^{a_{1}}_{n+1}\dots Q^{a_{N_{c}}}_{n+N_{c}} \rmd\tau^{a_{1}}\wedge\dots\wedge\tau^{a_{N_{c}}} 
  = \det\left(iQ^{a}_{n+b}\right)\rmd^{N_{c}}\tau\,,
\end{equation}
where the determinant is over the $a$ and $b$ indices. The partition function may then be written as
\begin{equation}
\label{PF_CI1}
  \cZ = i^{N_{f}}\frac{\det (Q^{a}_{n+b})^{2}}{(2\pi)^{N_{c}}}\int \rmd^{n}x\wedge\rmd^{n}\bar{x}\wedge\rmd^{m}p\wedge\rmd^{m}\bar{p}\,e^{W-\overline{W}} 
    \int \rmd^{N_{c}}\tau\wedge\rmd^{N_{c}}\bar{\tau}\det(M^{\dagger}M)\prod_{a}\delta(2\mu_{a}+\chi_{a})
\end{equation}
where the moment map depends on the imaginary part of $\tau^{a}$ according to
\begin{equation}
  -2\mu_{a} = \sum_{I=1}^{n} e^{-2\sum_{b}Q^{b}_{I}\tau^{b}_{2}} Q^{a}_{I}|x_{I}|^{2} + \sum_{I=n+1}^{n+N_{c}} Q^{a}_{I}e^{-2\sum_{b}Q^{b}_{I}\tau^{b}_{2}} 
  + \sum_{\alpha=N_{f}-m+1}^{N_{f}} Q^{a}_{\alpha}e^{-2\sum_{b}Q^{b}_{\alpha}\tau^{b}_{2}}|p_{\alpha}|^{2}\,,
\end{equation}
and the mass matrix $M^{\dagger}M$ can be expressed in terms of the moment maps via
\begin{equation}
  (M^{\dagger}M)_{ab}= \frac{\partial\mu_{a}}{\partial \tau^{b}_{2}}\,.
\end{equation}
This last relation implies that $\det(M^{\dagger}M)$ is precisely the inverse of the Jacobian factor produced by the coordinate transformation $\{\tau_{2}^{a}\}\rightarrow\{\mu_{a}\}$.
Consequently, the integration over the space of complex gauge orbits can be carried out leading to the numerical factor
\begin{equation}
  \int \rmd^{N_{c}}\tau\wedge\rmd^{N_{c}}\bar{\tau}\det(M^{\dagger}M)\prod_{a}\delta(2\mu_{a}+\chi_{a}) = (-2i\pi)^{N_{c}}\,.
\end{equation}
With the space of gauge orbits integrated out, the partition function \eqref{PF_CI1} assumes the simple form
\begin{equation}
\label{PF_CI2}
  \cZ =  i^{n+m} \det(Q^{a}_{n+b})^{2}\int \rmd^{n}x\wedge \rmd^{n}\bar{x} \wedge \rmd^{m}p\wedge\rmd^{m}\bar{p}\ e^{\sum_{\alpha}(p_{\alpha}G_{\alpha}-\bar{p}_{\alpha}\bar{G}_{\alpha})}\,.
\end{equation}
As in the last two examples, the $p$ integrals impose the embedding equation constraints $G_{\alpha}=0$ which can be used to solve for $m$ of the coordinates $x_{I}$. This leads to the partition function
\begin{equation}
\label{PF_CI}
  \cZ = i^{n-m} \frac{\det(Q^{a}_{n+b})^{2}}{(2\pi)^{2m}} \sum_{ \substack{ \{x_{n-m+\beta}|G_{\alpha}=0\} \\ \{\bar{x}_{n-m+\beta}|\bar{G}_{\alpha}=0\} } } 
  \int \frac{ \rmd x_{1}\wedge \dots \wedge \rmd x_{n-m}}{\det\left(\partial_{n-m+\beta} G_{\alpha}(x)\right)}
  \wedge \frac{ \rmd \bar{x}_{1}\wedge \dots \wedge \rmd \bar{x}_{n-m}}{\det\left(\bar{\partial}_{n-m+\beta} \bar{G}_{\alpha}(\bar{x})\right)}\,.
\end{equation}
The resulting nowhere vanishing holomorphic $n-m$ form $\Omega$ is given by
\begin{equation}
  \Omega = \frac{\det(Q^{a}_{n+b})}{(2\pi)^{m}}\frac{ \rmd x_{1}\wedge \dots \wedge \rmd x_{n-m}}{\det\left(\partial_{n-m+\beta} G_{\alpha}(x)\right)}\,,
\end{equation}
where the determinant in the denominator is over the $\alpha$ and $\beta$ indices, thus realizing from gauge theory the formulae for the holomorphic form on a Calabi-Yau in \cite{Strominger:1985it,Candelas:1987kf,Berglund:1993ax}.

\section{Discussion}
\label{sec:disc}

In this paper we have computed the exact partition function of abelian two dimensional $\cN=(2,2)$ gauge theories   capturing the K\"ahler potential in the complex structure moduli space of Calabi-Yau manifolds 
\rf{complex}\rf{formulacplex}. These path integrals can be enriched by the  insertion  of operators in the chiral ring and their conjugates, which   compute  the metric in the moduli space. It would be interesting 
to extend the computations in this paper by considering the   hemisphere function and to compute the associated D-brane charges, as well as to probe the geometric interpretation of conformal defects in the Calabi-Yau NLSM.

 The results of this paper, combined with those in \cite{Doroud:2012xw,Benini:2012ui},  give the exact answer to the inequivalent  supersymmetric partition functions on the two-sphere. 
While the metric in the moduli space is unambiguously defined, the two-sphere partition functions suffer  from some ambiguities, and can be thought of as  generating functions for the 
metrics on K\"ahler and complex structure moduli space.\footnote{The more invariant quantity is the product of the two partition functions $\cZ_{A}\cdot \cZ_{B}$.}
In particular, the partition function has an ambiguity associated to turning on an arbitrary twisted superpotential for the 
complex structure moduli, and this modifies the result of the partition function.

The computation of the K\"ahler potentials \cite{Jockers:2012dk,Gomis:2012wy} in both the K\"ahler moduli space and the complex structure moduli space are achieved by a direct gauge theory computation. 
This GLSM approach does not rely on mirror symmetry, and in fact has been used in \cite{Jockers:2012dk} to compute the K\"ahler potential for Calabi-Yau manifolds for which a geometrical mirror is not currently known.

It is nevertheless of interest to try to use  all these results to  find   new mirror manifolds. A geometric mirror manifold is expected to exist whenever a  Calabi-Yau manifold has a point in its moduli space of maximal 
unipotent monodromy. Mirror symmetry maps the K\"ahler potential in K\"ahler moduli space of  a Calabi-Yau $M$ to the K\"ahler potential in complex structure  moduli space of  the mirror Calabi-Yau $W$, and viceversa. 
In terms of the various two-sphere partition functions, mirror symmetry implies   that 
\beq
\cZ_{A}(M)=\cZ_{B}(W)\qquad \cZ_{B}(M)=\cZ_{A}(W)\,.
\eeq
The  existence of explicit formulae for these partition functions provides a window in which this problem can be addressed.

\section*{Acknowledgements}

We would like to thank  Davide Gaiotto, Kentaro Hori, Bruno Le Floch, Sungjay Lee, Joshua Lapan and  Martin Ro\v{c}ek  for discussions. This research was supported in part by Perimeter Institute for Theoretical Physics. Research at Perimeter Institute is supported 
by the Government of Canada through Industry Canada and by the Province of Ontario through the Ministry of Research and Innovation.
J.G. also acknowledges further support from an NSERC Discovery Grant and from an ERA grant by the Province of Ontario.

\clearpage


\appendix
\addtocontents{toc}{\protect\setcounter{tocdepth}{1}}

\section{Supersymmetry Algebra on \texorpdfstring{$S^{2}$}{S2} }
\label{app:susy}

The superconformal algebra in the canonical basis on $S^{2}$ was written down in \cite{Doroud:2012xw}. Here we present explicitly the  map between the $S^{2}$ basis and the standard Virasoro basis and extract the
two distinct $SU(2|1)$ subalgebras. The map between the two basis was also presented in \cite{Hori:2013ika}.

\subsection{The Superconformal Algebra in the Standard Basis}

The globally defined $\cN=(2,2)$ superconformal group in two dimensions is generated by the bosonic symmetries $\{J_{0},L_{0},L_{\pm};\bar{J}_{0},\bar{L}_{0},\bar{L}_{\pm}\}$ and the fermionic generators
$\{G^{\pm}_{\pm};\bar{G}^{\pm}_{\pm}\}$ satisfying the (anti-) commutation relations \cite{Polchinski:1998rr}
\begin{equation}
\label{SCA_Polchinski}
\begin{aligned}
  \left[L_{0},G^{s}_{\pm}\right] &= \mp \frac{1}{2} G^{s}_{\pm}
  \\
  \left[L_{\pm},G^{s}_{\mp}\right] &= \pm G^{s}_{\pm}
  \\
  \left[J_{0},G^{\pm}_{s}\right] &= \pm  G^{\pm}_{s}
  \\
  \left[L_{m},L_{n}\right] &= (m-n) L_{m+n}
  \\
  \left\{ G^{+}_{\pm}, G^{-}_{\pm} \right\} &= 2 L_{\pm}
  \\
  \left\{ G^{+}_{\pm}, G^{-}_{\mp} \right\} &= 2 L_{0} \pm J_{0}
\end{aligned}
\hspace{50pt}
\begin{aligned}
  \left[\bar{L}_{0},\bar{G}^{s}_{\pm}\right] &= \mp \frac{1}{2} \bar{G}^{s}_{\pm}
  \\
  \left[\bar{L}_{\pm},\bar{G}^{s}_{\mp}\right] &= \pm \bar{G}^{s}_{\pm}
  \\
  \left[\bar{J}_{0},\bar{G}^{\pm}_{s}\right] &= \pm  \bar{G}^{\pm}_{s}
  \\
  \left[\bar{L}_{m},\bar{L}_{n}\right] &= (m-n) \bar{L}_{m+n}
  \\
  \left\{ \bar{G}^{+}_{\pm}, \bar{G}^{-}_{\pm} \right\} &= 2 \bar{L}_{\pm}
  \\
  \left\{ \bar{G}^{+}_{\pm}, \bar{G}^{-}_{\mp} \right\} &= 2 \bar{L}_{0} \pm \bar{J}_{0}
\end{aligned}
\end{equation}
with all the other (anti-) commutations vanishing.
This algebra admits an automorphism $\sigma$ whose action on the generators is given by
\begin{equation}
\label{ABAutormorphism}
  \sigma\left(G^{\pm}_{\pm}\right)=G^{\mp}_{\pm},\quad \sigma(J_{0})=-J_{0}, \quad \sigma=1\ \text{otherwise}.
\end{equation}
We shall see below that this is precisely the map between the A-type and the B-type subalgebras.

\subsection{The Superconformal Algebra in the $S^{2}$ Basis}

The $\cN=(2,2)$ superconformal algebra in the $S^{2}$ basis is spanned by the bosonic generators
\begin{equation}
J_{m}, K_{m}, R, \cA\,,
\end{equation}
and the supercharges
\begin{equation}
Q_{\alpha}, S_{\alpha}, \bar{Q}_{\alpha}, \bar{S}_{\alpha}\,.
\end{equation}
$J_m$ generate the $SU(2)$ isometries of $S^2$ while $K_m$ generate the conformal symmetries of $S^2$.
$R$ and  $\cA$ are each a $U(1)$ $R$-symmetry generator, the first being non-chiral and the latter being chiral.
These generators are related to the standard generators introduced above via
\begin{equation}
\label{mapSCAPtoSCAS2}
\begin{aligned}
  J_{1} &= \frac{i}{2} \left( L_{-} + L_{+} + \bar{L}_{-} + \bar{L}_{+} \right)
  \\
  J_{2} &= \frac{1}{2} \left( L_{-} - L_{+} - \bar{L}_{-} + \bar{L}_{+} \right)
  \\
  J_{3} &= L_{0} - \bar{L}_{0}
  \\
  R &= J_{0} + \bar{J}_{0}
  \\
  S &= \frac{1}{\sqrt{2}} \left( \begin{array}{c}
				  G^{+}_{+} + i \bar{G}^{+}_{-}
				  \\
				  iG^{+}_{-} + \bar{G}^{+}_{+}
                                 \end{array}
                          \right)
  \\
  Q &= \frac{1}{\sqrt{2}} \left( \begin{array}{c}
				  -iG^{-}_{+} - \bar{G}^{-}_{-}
				  \\
				  G^{-}_{-} + i \bar{G}^{-}_{+}
                                 \end{array}
                          \right)
\end{aligned}
\hspace{50pt}
\begin{aligned}
  K_{1} &= -\frac{1}{2} \left( L_{-} + L_{+} - \bar{L}_{-} - \bar{L}_{+} \right)
  \\
  K_{2} &= \frac{i}{2} \left( L_{-} - L_{+} + \bar{L}_{-} - \bar{L}_{+} \right)
  \\
  K_{3} &= i\left( L_{0} + \bar{L}_{0} \right)
  \\
  \cA &= - J_{0} + \bar{J}_{0}
  \\
  \bar{S} &=\frac{1}{\sqrt{2}} \left( \begin{array}{c}
					G^{-}_{+} + i \bar{G}^{-}_{-}
					\\
					iG^{-}_{-} + \bar{G}^{-}_{+}
				      \end{array}
			       \right)
  \\
  \bar{Q} &=\frac{1}{\sqrt{2}} \left( \begin{array}{c}
					iG^{+}_{+} + \bar{G}^{+}_{-}
					\\
					-G^{+}_{-} -i \bar{G}^{+}_{+}
				      \end{array}
			       \right)
\end{aligned}
\end{equation}
and satisfy the algebra\footnote{The generator of the $U(1)$ axial symmetry $\cA$ used here defers from the one used in \cite{Doroud:2012xw} by a factor of $i$.}
\begin{equation}
\label{Neq2sca}
\begin{aligned}
  \{ S_{\alpha}, Q_{\beta} \} &= \gamma^{m}_{\alpha \beta} J_{m} - \frac{1}{2} C_ {\alpha \beta}R \quad &
  [J_{m},S^{\alpha}] &= -\frac{1}{2} \gamma_{m}^{\ \alpha \beta}S_{\beta}\quad &
  [R,S_{\alpha}] &= + S_{\alpha}
\\ \{ \bar{S}_{\alpha}, \bar{Q}_{\beta} \} &= -\gamma^{m}_{\alpha \beta} J_{m} - \frac{1}{2} C_{\alpha \beta}R&
  [J_{m},Q^{\alpha}] &= -\frac{1}{2} \gamma_{m}^{\ \alpha \beta}Q_{\beta} &
  [R,Q_{\alpha}] &= - Q_{\alpha}
\\
 \{ Q_{\alpha}, \bar{Q}_{\beta} \} &= \gamma^{m}_{\alpha \beta} K_{m} - \frac{i}{2} C_{\alpha \beta} \cA  &
  [J_{m},\bar{Q}^{\alpha}] &= -\frac{1}{2} \gamma_{m}^{\ \alpha \beta}\bar{Q}_{\beta} &
  [R,\bar{Q}_{\alpha}] &= + \bar{Q}_{\alpha}
\\
      \{ S_{\alpha}, \bar{S}_{\beta} \} &= \gamma^{m}_{\alpha \beta} K_{m} + \frac{i}{2} C_{\alpha \beta} \cA \quad  &
  [J_{m},\bar{S}^{\alpha}] &= -\frac{1}{2} \gamma_{m}^{\ \alpha \beta}\bar{S}_{\beta} &
  [R,\bar{S}_{\alpha}] &= - \bar{S}_{\alpha}
\\
  [J_{m},J_{n}] &= i \epsilon_{mnp} J^{p}\quad  &
  [K_{m},S^{\alpha}] &= -\frac{1}{2} \gamma_{m}^{\ \alpha \beta}\bar{Q}_{\beta} &
  [\cA, S_{\alpha}] &= i\bar{Q}_{\alpha}
\\
  [K_{m},K_{n}] &= -i \epsilon_{mnp} J^{p} &
  [K_{m},Q^{\alpha}] &= -\frac{1}{2} \gamma_{m}^{\ \alpha \beta}\bar{S}_{\beta} &
  [\cA, Q_{\alpha}] &= - i\bar{S}_{\alpha}
\\
 [J_{m},K_{n}] &= i \epsilon_{mnp} K^{p}	 &
  [K_{m},\bar{Q}^{\alpha}] &= -\frac{1}{2} \gamma_{m}^{\ \alpha \beta}S_{\beta} &
  [\cA, \bar{Q}_{\alpha}] &= -i S_{\alpha}
\\
  & &
  [K_{m},\bar{S}^{\alpha}] &= -\frac{1}{2} \gamma_{m}^{\ \alpha \beta}Q_{\beta} &
  [\cA, \bar{S}_{\alpha}] &= i Q_{\alpha}\,.
\end{aligned}
\end{equation}

This algebra admits a $\mathbb{Z}_2$ automorphism, under which
\begin{equation}
\begin{aligned}
J_m, R, Q_\alpha,S_\alpha&\rightarrow J_m, R, Q_\alpha,S_\alpha\\
K_m, \cA, \bar Q_\alpha,\bar S_\alpha&\rightarrow -K_m,-\cA,- \bar Q_\alpha,-\bar S_\alpha\,.
\end{aligned}
\end{equation}
The generators $\{J_{m},R,S,Q\}$ form a subalgebra which is the A-type $SU(2|1)$ algebra used in \cite{Doroud:2012xw}.
In addition to this automorphism, the algebra \eqref{Neq2sca} inherits the automorphism $\sigma$ defined in \eqref{ABAutormorphism}.
This implies that $\{\sigma(J_{m}), \sigma(R), \sigma(S), \sigma(Q)\}$ given by
\begin{equation}
\begin{aligned}
  \sigma(J_{m}) &= J_{m}
  \\[10pt]
  \sigma(R) &= \cA
\end{aligned}
\hspace{50pt}
\begin{aligned}
  \sigma(S) &= \frac{S + \bar{S}}{2} + i\frac{Q + \bar{Q}}{2}
  \\
  \sigma(Q) &= -i\frac{S - \bar{S}}{2} +\frac{Q - \bar{Q}}{2}
\end{aligned}
\end{equation}
also form a $SU(2|1)$ subalgebra with the (anti-) commutation relations
\begin{equation}
\begin{aligned}
 \ [J_{m},J_{n}] &= i \epsilon_{mnp} J_p~
& [J_{m},\sigma(Q)] &= -\frac{1}{2} \gamma_{m} \sigma(Q)\qquad
  &  [J_{m},\sigma(S)] &=-\frac{1}{2} \gamma_{m}\sigma(S)
\\
\{\sigma(S)_{\alpha},\sigma(Q)_{\beta}\} &= \gamma^{m}_{\alpha \beta} J_{m} - \frac{1}{2} C_ {\alpha\beta}\cA~ &
[\cA,\sigma(Q)] &= - \sigma(Q) \qquad &
  [\cA,\sigma(S)] &=  \sigma(S)\,,
\end{aligned}
\label{SU(2|1)_B}
\end{equation}
which is precisely the B-type algebra we sought. One can check that the realization of this algebra on the vector and chiral multiplets matches the realization of the A-type $SU(2|1)$ algebra
on the twisted vector and twisted chiral multiplets up to field redefinitions.

\subsection{Choice of Localizing Supercharge}

Our choice of localizing supercharge for the theory of twisted vector and twisted chiral multiplets is $\cQ=S_{1}+Q_{2}$. The corresponding choice of killing spinors is given by \eqref{cQKS} and the parameters of 
the $U(1)\times U(1)_R\times G$ transformations generated by $\cQ^2$ are completely determined through \rf{spinorbilinears}\rf{spinorbilinearsb}\rf{gaugetransf}. The Killing vector   \rf{spinorbilinears}
\beq
 v^{\mu}\partial_{\mu} = i \bar{\epsilon}\gamma^{\mu}\epsilon \partial_{\mu} = \frac{1}{r} \partial_{\varphi}
\eeq
has indeed fixed points at the poles of the two-sphere. The $U(1)_R$ transformation parameter \rf{spinorbilinearsb} is  given in terms of the radius of the two-sphere
\beq
\alpha=-\frac{1}{2r} \bar{\epsilon}\gamma^{\hat 3}\epsilon = - \frac{1}{2r}\,.
\eeq
The induced gauge transformations \rf{gaugetransf} are
\beq
\Lambda=-\frac{1}{2}\sin\theta \, e^{ +i \varphi}
\sigma\qquad\quad \bar\Lambda=  \frac{1}{2}\sin\theta \, e^{ -i \varphi}  \bar{\sigma}\qquad \quad \Omega=- \frac{1}{r} A_{\varphi}\,.
\eeq
The gauge transformation parameters are invariant under $\cQ^2=J_3+\frac{R}{2}$.

\section{BRST Supercharge and Gauge Fixing}
\label{app:BRST}

As in any gauge theory, the formalism we have used has built in it a large redundancy which we need to remove in order to proceed with our computation of the partition function. This is achieved by introducing the 
supercharge $\cQ_{\text{BRST}}$ and the ghost and anti-ghost multiplets $\{c,a_{\circ}\}$ and $\{\bar{c},b\}$, where $c$ and $\bar{c}$ are Grassmann odd and $a_{\circ}$ and $b$ are Grassmann even scalars and they 
all have vanishing R-charge.

In terms of the ghost multiplet fields, the BRST operator is defined as
\begin{equation}
\label{BRST}
  \cQ_{\text{BRST}} = \delta_{G}(c) , 
  \hspace{40pt} 
  \cQ_{\text{BRST}}^{2} = \delta_{G}(a_{\circ})\,,
\end{equation}
where $a_{\circ}$ is assumed to be supersymmetric \emph{i.e.} $\cQ a_{\circ}=0$. By construction, adding the BRST supercharge to the supersymmetry algebra \eqref{su11} leaves the algebra 
invariant up to gauge transformations. We therefore define the supercharge $\hat{\cQ} = \cQ + \cQ_{\text{BRST}}$ and require that it realizes the $su(1|1)$ algebra \eqref{su11} as
\begin{equation}
\label{GFsu11}
  \hat{\cQ}^{2} = \cL_{v} - \frac{i}{2r} R + \delta_{G}(a_{\circ})
\end{equation}
where $\cL_{v}$ denotes the Lie(-Lorenz) derivative along $v = 1/r\partial_{\varphi}$ and $R$ is the generator of the $U(1)_{R}$ symmetry. This fixes the supersymmetry transformation rule for
the ghost and anti-ghost multiplet fields completely. The action of $\hat{\cQ}$ on the ghost multiplet fields is found to be
\begin{equation}
  \hat{\cQ} c = a + i \, cc + v^{\mu}A_{\mu} + \frac{1}{2} \sin\theta \left( \sigma e^{+i\varphi} - \bar{\sigma} e^{-i \varphi} \right), 
  \hspace{40pt} 
  \hat{\cQ}a_{\circ} = i [c,a_{\circ}]\,,
\end{equation}
while the anti-ghost multiplet fields transform as
\begin{equation}
  \hat{\cQ} \bar{c} = i b,
  \hspace{40pt}
  \hat{\cQ} b = -i (\cL_{v} + i [a,\cdot\,]) \bar{c} \,.
\end{equation}
We remark that by construction the action of $\hat{\cQ}$ and $\cQ$ coincide on all gauge invariant objects. In particular the deformation term \eqref{V} satisfies $\hat{\cQ}\cV=\cQ\cV$.

For a choice of gauge fixing functional $\cG[A,\Phi]$, the gauge fixing condition, $\cG=0$, can be imposed on the path integral in a supersymmetric way by adding the deformation term
$\hat{\cQ} \cV_{\text{G.F.}}$ to the action where
\begin{equation}
\label{GFV}
  \cV_{\text{G.F.}} = \frac{1}{2} \int \rmd^{2}x \sqrt{h} \, \Tr \left\{ \bar{c} \left(\cG  -\frac{i}{4}b\right) \right\}\,.
\end{equation}
Being exact in $\hat{\cQ}$, this choice of deformation term guarantees the independence of the path integral from the choice of gauge fixing functional $\cG$, provided that the ghost kinetic term, 
$\bar{c}\cQ_{\text{BRST}}\cG$, is non-degenerate.

In the presence of a Higgs branch, such as the saddle points \eqref{higgsbranch}, a particularly convenient choice for the gauge fixing functional $\cG$ is the so called $R_{\xi}$ gauge (with $\xi=1$)
\begin{equation}
\label{R_xi}
  \cG = \nabla_{\mu}A^{\mu} + i \left(Y \bar{Y}_{\circ} - Y_{\circ}\bar{Y} \right)
\end{equation}
We remark that the gauge fixing condition on the saddle points reduces to the usual Lorenz gauge $\nabla_{\mu}A^{\mu}=0$ which is compatible with the choice $A_{\mu}=0$ in \eqref{higgsbranch}.
\section{\texorpdfstring{$\hat{\cQ}$}{Q}-Exact Deformation Term}
\label{app:QV}

Here we spell out the precise deformation term $\hat{\cQ}\cV^{\prime}$, including all the total derivative terms, which we use for the localization computation. We break $\cV^{\prime}$ into four pieces corresponding
to the twisted vector, twisted chiral, Fayet-Iliopoulos and gauge fixing terms
\begin{equation}
  \cV^{\prime} =  \cV_{\text{t.v.m.}} + \cV_{\text{G.F.}} + \cV_{\text{F.I.}} + \sum_{I} \cV^{I}_{\text{t.c.m.}}\,.
\end{equation}
For concreteness, let $\{T_{a},\, a=1,\dots,\dim \mathfrak{g}\}$ be the set of normalized generators of the gauge algebra $\mathfrak{g}$. The twisted vector multiplet as well as the ghost fields are valued
in the adjoint representation of $\mathfrak{g}$ while the twisted chiral multiplet fields live in a representation $\mathbf{r}$ of $\mathfrak{g}$. Suppressing the integration over the sphere, the various terms
in $\cV^{\prime}$ are given by
\begin{equation}
\label{fullV}
\begin{aligned}
  \cV_{\text{t.v.m.}} 
		&= \frac{1}{4}\Tr\left\{
			\bar{\epsilon}\gamma^{\hat 3}\bar{\eta} \left(\rmD + i F\right) + \epsilon\gamma^{\hat 3}\eta \left(\rmD - i F\right)
			+ \frac{i}{2} \left( \bar{\epsilon} \eta + \epsilon \bar{\eta} \right) [\sigma ,\,  \bar{\sigma} ] 
			- i \bar{\epsilon}\gamma^{\hat 3} \slashed{D} \bar{\sigma} \, \eta - i \epsilon\gamma^{\hat 3}\slashed{D}\sigma \, \bar{\eta}
			\right\}\,,
  \\
  \cV_{\text{G.F.}} &= \frac{1}{2} \Tr \left\{ \bar{c} \left(\nabla_{\mu}A^{\mu} + i g^{2} \left(Y \bar{Y}_{\circ} - Y_{\circ}\bar{Y} \right)  -\frac{i}{4}b\right) \right\}\,,
\\
  \cV_{\text{F.I.}} &= \frac{\chi}{2i}\Tr \left( \bar{\epsilon}\gamma^{\hat 3}\bar{\eta} + \epsilon\gamma^{\hat 3}\eta \right)\,,
  \\
  \cV_{\text{t.c.m.}} 
		&= \frac{1}{2} \Big[ \bar{G} \left( \bar{\epsilon}\zeta_{-} + \epsilon\zeta_{+} \right) - \left( \bar{\epsilon}\bar{\zeta}_{+} + \epsilon\bar{\zeta}_{-} \right) G
		   +i\bar{Y} D_{\mu}\left( \bar{\epsilon}\gamma^{\mu}\zeta_{-} + \epsilon\gamma^{\mu}\zeta_{+} \right)
		   -i \bar{Y} \left( \bar{\sigma} \bar{\epsilon} \zeta_{+} + \sigma \epsilon \zeta_{-} \right)
		   \\[3pt]
		   & \qquad\
		   -i D_{\mu}\left( \bar{\epsilon}\gamma^{\mu}\bar{\zeta}_{+} + \epsilon\gamma^{\mu}\bar{\zeta}_{-}  \right) Y
		   +i \left( \bar{\epsilon}\bar{\zeta}_{-} \bar{\sigma} + \epsilon\bar{\zeta}_{+} \sigma \right) Y 
		   +i \bar{Y} \left( \bar{\epsilon}\gamma^{\hat 3}\bar{\eta} + \epsilon\gamma^{\hat 3}\eta \right) Y
		   \Big]\,.
\end{aligned}
\end{equation}
where there is an independent Fayet-Iliopoulos  parameter $\chi_{a}$ for each $U(1)$ factor in the gauge group.\footnote{Without loss of generality, we have not chosen an independent parameter for all the different
$\hat{\cQ}$-exact pieces in the deformation term since $\hat{\cQ}$-exactness guarantees that the final result will be independent of such parameters.}
As we alluded to in section \ref{sec:localization}, the twisted vector and twisted chiral terms may be written in a more compact form as
\begin{equation}
\begin{aligned}
  \cV_{\text{t.v.m.}} 
 		&= \frac{1}{4} \widetilde{\cQ} \Tr \left( \eta \gamma^{\hat 3} \bar{\eta} + \frac{i}{r} \sigma \bar{\sigma} \right)\,,
 		\\
   \cV_{\text{t.c.m.}} 
 		&= \frac{1}{2} \widetilde{\cQ} \left(\bar{G}Y - \bar{Y}G + \frac{i}{r}\bar{Y}Y \right)\,,
\end{aligned}
\end{equation}
where $\widetilde{\cQ}=S_{1}-Q_{2}$. Using \eqref{fullV}, the deformation term may be split into bosonic and fermionic pieces, up to a total derivative term, \emph{i.e.}
\begin{equation}
\label{app:tQV}
  \hat{\cQ}\cV = \hat{\cQ}\cV\big|_{\text{bos.}} + \hat{\cQ}\cV\big|_{\text{fer.}} + \nabla_{\mu}J^{\mu}
\end{equation}
where the bosonic part is given by
\begin{equation}
\label{tQVbos}
\begin{aligned}
  t\hat{\cQ}\cV\big|_{\text{bos.}} &=
  	\frac{t}{2} \sum_{a}\left\{ F_{a}^{2} + (D^{\mu}\sigma)_{a} (D_{\mu}\bar{\sigma})_{a} + \frac{1}{4} [ \sigma,\,\bar{\sigma}]_{a}^{2} +\widetilde{\rmD}_{a}^{2} + \widetilde{b}_{a}^{2} + \cG_{a}^{2} 
	+ \left( \bar{Y}T_{a}Y - \chi \right)^{2}
	\right\}
	\\
	&\quad
	+ t \left( \bar{G} G + D^{\mu}\bar{Y} D_{\mu}Y + \frac{1}{2} \bar{Y}\left\{\sigma,\, \bar{\sigma}\right\} Y \right)
\end{aligned}
\end{equation}
with $\widetilde{\rmD}_{a} = \rmD_{a} + i (\bar{Y}T_{a}Y - \chi )$ and $\widetilde{b}= b/2+i\cG$. The fermionic part of $\hat{\cQ}\cV$ is given by
\begin{equation}
\label{tQVfer}
\begin{aligned}
  t\hat{\cQ}\cV\big|_{\text{fer.}} &=
  	-\frac{it}{2} \Tr\left\{
		\bar{\eta}\left(\slashed{D} + \frac{1}{r}\gamma^{\hat 3} \right) \eta + \sigma\, \bar{\eta}\gamma^{\hat 3}\bar{\eta} -\bar{\sigma}\, \eta\gamma^{\hat 3}\eta - i \bar{c} \hat{\cQ}\cG 
		+ \frac{i}{4}\bar{c}\left(v^{\mu}\partial_{\mu} + i [a_{\circ},\,\cdot\,] \right) \bar{c}
		\right\}
		\\
		&\quad
		+it \left( 
		\bar{Y}(\bar{\eta}_{-} - \eta_{+}) \zeta + \bar{\zeta}(\bar{\eta}_{+} - \eta_{-})Y +\bar{\zeta} \left( \slashed{D} - \sigma \gamma_{-} - \bar{\sigma} \gamma_{+} \right)\zeta
		\right)
\end{aligned}
\end{equation}
and the total derivative term may be written as
\begin{equation}
\begin{aligned}
  J^{\mu} &=  \frac{i}{2} \Tr\left\{
  		(\bar{\epsilon} \epsilon) \varepsilon^{\mu\nu} \bar{\sigma} D_{\nu} \sigma + \frac{1}{2r} v^{\mu} \bar{\sigma} \sigma  
		+ \frac{1}{2} \left( \bar{\epsilon}\gamma^{\mu}\bar{\eta} \right)\left( \epsilon\gamma^{\hat 3}\eta \right) - \frac{1}{2} \left( \epsilon\bar{\eta} \right)\left( \bar{\epsilon} \gamma^{\hat 3}\gamma^{\mu} \eta \right)
		\right\}
		\\
		&\quad
		+\sum_{I} \left( 
		(\bar{\epsilon}\gamma_{-}\epsilon) D^{\mu}\bar{Y}_{I}Y_{I} - (\bar{\epsilon}\gamma_{+}\epsilon) \bar{Y}_{I}D^{\mu}Y_{I} - \frac{i}{2} \left( \bar{\epsilon}\gamma^{\hat 3}\gamma^{\mu}\epsilon\right)
		\left( \bar{G}_{I}Y_{I} + \bar{Y}_{I} G_{I} \right) + i (\bar{\epsilon}\gamma_{-}\epsilon)\left( \bar{\zeta}_{I} \gamma^{\mu} \zeta_{I} \right)
		\right)
		\\
		&\quad
		+\frac{\chi}{2} \Tr\left\{
		\left( \bar{\epsilon}\gamma^{\hat 3}\gamma^{\mu}\bar{\epsilon} \right) \bar{\sigma}+\left( \epsilon\gamma^{\hat 3}\gamma^{\mu}\epsilon \right)\sigma
		\right\}\,.
\end{aligned}
\end{equation}

\section{One-Loop Determinant}
\label{app:1-loop}

Consider the Abelian gauge theory with gauge group $U(1)^{N_{c}}$, minimally coupled to $N_{f}$ twisted chiral multiplets with generic charges $\{Q^{a}_{I}| a=1,\dots,N_{c} ; I=1,\dots,N_{f}\}$. We 
assume $N_{f}\ge N_{c}$ since, as will become clear, the one-loop determinant vanishes for $N_{c}>N_{f}$ due to fermionic zero modes.

Deforming the path integral by adding the deformation term $t\hat{\cQ}\cV$ to the action
and taking the large $t$ limit, the path integral localizes to the saddlepoints which are constant maps subject to the $\rmD$-term constraints
\begin{equation}
\label{saddles}
  \Big\{ Y=\text{constant} \, \Big| \,
  \sum_{I} Q^{a}_{I} |Y_{I}|^{2} = \chi^{a}  \Big\}\,.
\end{equation}
The measure of integration is defined by the one-loop -- with respect to $t$ -- fluctuations of the fields around these saddle points. To extract this measure, we expand $\hat{\cQ}\cV$ to quadratic order
around the saddle points \eqref{saddles}, we therefore redefine the fields as
\begin{equation}
  \Phi \rightarrow \frac{1}{\sqrt{t}} \Phi
\end{equation}
for twisted vector fields and
\begin{equation}
Y_{I}\rightarrow Y_{I} + \frac{1}{\sqrt{t}} y_{I}\,,\quad G_{I} \rightarrow\frac{1}{\sqrt{t}}G_{I}\,,\quad \zeta_{I} \rightarrow \frac{1}{\sqrt{t}}\zeta
\end{equation}
for twisted chiral fields.
Imposing the $R_{\xi}$ gauge on the gauge field fluctuations
\begin{equation}
  \cG_{a} = \frac{1}{\sqrt{t}} \left( \nabla_{\mu} A^{\mu}_{a} - i \sum_{I} Q^{a}_{I} \left(\bar{y}_{I}Y_{I} - \bar{Y}_{I}y_{I}\right)\right)
\end{equation}
the quadratic part of the deformation term can be cast into the following form
\begin{equation}
\label{qvquad1}
\begin{aligned}
    t\hat{\cQ}\cV\big|_{\text{quad}} &=
    	\frac{1}{2 r^{2}}\sum_{a,b} \left[
	A^{\mu}_{a} \left(2M^{2}_{ab} + \delta_{ab} - r^{2} \delta_{ab} \nabla^{2} \right)A^{b}_{\mu}
	+ (\bar{\sigma}_{a}\,,\,\bar{c}_{a}) \left( 2M^{2}_{ab} - r^{2}\delta_{ab} \nabla^{2} \right) (\sigma_{b}\,,\,c_{b})^{T}
	\right]
	\\
	&\quad
	+\frac{1}{r^{2}}\sum_{I,J} \bar{y}_{I} \left(2 M^{2}_{IJ}-r^{2} \delta_{IJ}\nabla^{2} \right)y_{J}
	+i \sum_{a,I}Q_{I}^{a}\left[ \bar{\eta}_{a} \left(\bar{Y}_{I}\zeta_{+}^{I} + \bar{\zeta}_{-}^{I}Y_{I}  \right)  
	- \eta_{a} \left( \bar{Y}_{I}\zeta_{-}^{I} + \bar{\zeta}_{+}^{I}Y_{I} \right)\right]
	\\
	&\quad
	-\frac{i}{2r}\sum_{a}\bar{\eta}_{a}\left(r\slashed{\nabla}+\gamma^{\hat 3}\right)\eta_{a} + i \sum_{I} \bar{\zeta}_{I}\slashed{\nabla}\zeta_{I}
	+\sum_{I}\bar{G}_{I}G_{I} + \frac{1}{2}\sum_{a}\left( \widetilde{\rmD}^{2} + \widetilde{b}^{2}\right) + \sum_{a}\bar{c}_{a}K_{a}
\end{aligned}
\end{equation}
where $\bar{c}K$ summerizes the fermionic terms that do not contribute to the one-loop determinant. Explicitely, $K_{a}$ is given by
\begin{equation*}
      K_{a}=\frac{1}{8} v^{\mu}\partial_{\mu}\bar{c}_{a} - \frac{i}{4}\nabla_{\mu} \left(\epsilon\gamma^{\hat 3}\gamma^{\mu}\eta_{a}-\bar{\epsilon}\gamma^{\hat 3}\gamma^{\mu}\bar{\eta}_{a}\right)
	+\frac{i}{2}\sum_{I}Q^{a}_{I}\left( (\bar{\epsilon}\bar{\zeta}_{+}^{I}-\epsilon\bar{\zeta}_{-}^{I})Y_{\circ}^{I} - \bar{Y}_{\circ}^{I} (\bar{\epsilon}\zeta_{-}^{I} -\epsilon\zeta_{+}^{I}) \right)\,.
\end{equation*}
We define the $N_{f}\times N_{c}$ matrix $M$ and it's hermitian conjugate $M^{\dagger}$ as
\begin{equation}
  M^{\ a}_{I} = r Q^{a}_{I}Y_{I}\,,\qquad M^{\dagger\,I}_{a} = r Q^{a}_{I}\bar{Y}_{I}\,.
\end{equation}
The mass matrices $M^{2}_{ab}$ and $M^{2}_{IJ}$ appearing in \eqref{qvquad1} are then given by
\begin{equation}
  M^{2}_{ab}= (M^{\dagger}M)_{ab}\,,\qquad M^{2}_{IJ} = (MM^{\dagger})_{IJ}\,.
\end{equation}
For generic charges $Q^{a}_{I}$ and with $N_{f}\ge N_{c}$, both of these matrices are of rank $N_{c}$. Furthermore, one can easily check that they have the same eigenvalues since for any eigenvector $u$ of $MM^{\dagger}$, the
vector $M^{\dagger}u$ is an eigenvector of $M^{\dagger}M$ with the same eigenvalue.

From \eqref{qvquad1}, it is evident that the path integral over the auxiliary fields $\widetilde{\rmD}_{a},\widetilde{b}_{a}$ and $G_{I}$ is Gaussian and yields a trivial factor. It is also clear that the
path integration over the twisted vector scalars $\{\sigma,\bar{\sigma}\}$ and the ghost and anti-ghost fields $\{c,\bar{c}\}$ yield canceling contributions. 

As for the rest of the field, we begin our analysis by diagonalizing the Laplacian on the gauge field. Using the spectrum of the Laplacian operator on $T^{*}S^{2}$,
\begin{equation}
  \text{spectrum}(-r^{2}\nabla^{2}\big|_{T^{*}S^{2}}) = \{ {(J^{2}+J-1)}^{\times(4J+2)} ; J=1,2,\dots \}\,,
\end{equation}
we may compute the contribution of the gauge fields to the one-loop determinant to be
\begin{equation}
\label{1lA}
  \Delta_{A}^{-1} = \prod_{J=1}^{\infty}\det\left(\frac{J(J+1)}{2}\mathbbm{1}+2M^{\dagger}M\right)^{2J+1} \,.
\end{equation}
In order to compute the contribution to the one-loop determinant arising from the fluctuations of the twisted chiral scalar fields, we first need to isolate the zero modes
satisfying
\begin{equation}
\label{saddlepointfluctuations}
  \nabla_{\mu}y_{I}=0\quad\text{and}\quad \sum_{J} M^{2}_{IJ}\, y_{J} = 0\,.
\end{equation}
These are the longitudinal fluctuations that lie in the space of saddle points \eqref{saddles} and need to be excluded from the one-loop analysis. This amounts to removing the vanishing eigen values of $MM^{\dagger}$ from the
$J=0$ mode contribution. The contribution from the twisted chiral scalars is then
\begin{equation}
\label{1ly}
\begin{aligned}
  \Delta_{y}^{-1} &= \det{}^{\prime}(2MM^{\dagger})\prod_{J=1}^{\infty}\det\left(J(J+1)+2 M M^{\dagger}\right)^{2J+1}
  \\
  &= \det(2M^{\dagger}M)\ \prod_{J=1}^{\infty} \left[ J^{4J(N_{f}-N_{c})}\det\left(J(J+1)+2 M^{\dagger}M\right)^{2J+1} \right]\,.
\end{aligned}
\end{equation}
Putting \eqref{1lA} and \eqref{1ly} together, the boson and ghost contributions to the one-loop determinant takes the form
\begin{equation}
\label{1lb}
  \Delta_{b}^{-1} = \det(2M^{\dagger}M) \prod_{J=1}^{\infty}\left(\frac{1}{2}\right)^{(2J+1)N_{c}}\prod_{J=1}^{\infty} \left[J^{4J(N_{f}-N_{c})} \det\left(J(J+1)+2M^{\dagger}M\right)^{4J+2}\right]\,.
\end{equation}

To compute the contribution to the one-loop determinant due to fermionic fields, consider the field redefinition
\begin{equation}
\psi = \bar{\zeta}_{+} + \zeta_{-} \,,\qquad \bar{\psi} = \bar{\zeta}_{-} + \zeta_{+} \,.
\end{equation}
In terms of $\psi$ and $\bar{\psi}$, we may rewrite the quadratic fermion part of $t\hat{\cQ}\cV$ as
\begin{equation}
\label{Df}
  \bar{f}D_{f}f =
		\left(
		\begin{array}{c}
			\oplus_{a} \bar{\eta}_{a}  \\  \oplus_{I} \psi_{I}
		\end{array}
		\right)^{T}
		\left(
		\begin{array}{cc}
			\mathbbm{1} \otimes \left(-\frac{ir}{2} \slashed{\nabla} -\frac{i}{2} \gamma^{\hat 3}\right)
			& \quad
			+i (M^{\dagger}\otimes\gamma_{+}+M^{T}\otimes\gamma_{-})
			\\
			-i (M \otimes\gamma_{-} + M^{\ast}\otimes\gamma_{+}) 
			& 
			\mathbbm{1} \otimes ir\slashed{\nabla}
		\end{array}
		\right)
		\left(
		\begin{array}{c}
			\oplus_{a}\eta_{a}  \\  \oplus_{I} \bar{\psi}_{I}
		\end{array}
		\right)\,,
\end{equation}
where the operator $D_{f}$ is block diagonal, \emph{i.e.} it does not mix the eigenmodes of the Dirac operator. Exploiting this fact we may consider each block separately. In the $J$th mode, the Dirac operator
is diagonal while the chirality operator $\gamma^{\hat 3}$ has only non-zero off-diagonal elements. Explicitly, we have
\begin{equation}
  ir\slashed{\nabla}\big|_{J} = (J+1/2)\left(
				\begin{array}{cc}
				  1
				  &\ \
				  0
				  \\
				  0
				  &\,
				  -1
				\end{array}
				\right)\,,\qquad
  \gamma^{\hat 3}\big|_{J} = \left(
			\begin{array}{cc}
			  0
			  &\
			  1
			  \\
			  1
			  &\
			  0
			\end{array}
			\right)\,,
\end{equation}
in the basis of eigenspinors of the Dirac operator. In this basis, the $J$th block of the operator $D_{f}$ in \eqref{Df} takes the form
\begin{equation}
\label{DfJ}
  D_{f}[J]= \left(
		  \begin{array}{cc}
		    \mathbbm{1}\otimes
		    \left(
		    \begin{array}{cc}
		      -\frac{J+1/2}{2}
		      &\
		      -\frac{i}{2}
		      \\[3pt]
		      -\frac{i}{2}
		      &\
		      \frac{J+1/2}{2}
		    \end{array}
		    \right)
		    &\
		    \frac{i}{2}
		    \left(
		    \begin{array}{cc}
		      M^{\dagger}+M^{T}
		      &\
		      M^{\dagger}-M^{T}
		      \\
		      M^{\dagger}-M^{T}
		      &\
		      M^{\dagger}+M^{T}
		    \end{array}
		    \right)
		    \\
		    -\frac{i}{2}
		    \left(
		    \begin{array}{cc}
		      M^{\ast}+M
		      &\
		      M^{\ast}-M
		      \\
		      M^{\ast}-M
		      &\
		      M^{\ast}+M
		    \end{array}
		    \right)
		    &\
		    \mathbbm{1}\otimes
		    \left(
		    \begin{array}{cc}
		      J+\frac{1}{2}
		      &
		      0
		      \\
		      0
		      &
		      -J-\frac{1}{2}
		    \end{array}
		    \right)
		  \end{array}
		  \right)\,,
\end{equation}
and the fermion contribution to the one-loop determinant takes the form
\begin{equation}
\label{deltaf1}
 \Delta_{f} = \prod_{J=1/2}^{\infty} \big|D_{f}[J]\big|^{2J+1} = \prod_{J=1}^{\infty} \big|D_{f}[J-1/2]\big|^{2J}\,.
\end{equation}
The finite dimensional determinant $\big|D_{f}[J-1/2]\big|$ can easily be computed since the bottom right $N_{f}\times N_{f}$ block of \eqref{DfJ} is diagonal which allows us to put the matrix $D_{f}[J]$ in
a lower triangular form. This is achieved via the non-degenerate matrix
\begin{equation}
  U[J-1/2] = 
      \left(
      \begin{array}{cc}
	\mathbbm{1}
	&\
	-\frac{i}{2}
	 \left(
	 \begin{array}{cc}
	  M^{\dagger}+M^{T}
	  &\
	  M^{\dagger}-M^{T}
	  \\
	  M^{\dagger}-M^{T}
	  &\
	  M^{\dagger}+M^{T}
	\end{array}
	\right)
	\\
	0
	&\
	\mathbbm{1}
      \end{array}
      \right)\
      \left(
      \begin{array}{cc}
	\mathbbm{1}
	&\
	0
	\\
	0
	&\
	\mathbbm{1}\otimes
	  \left(
	  \begin{array}{cc}
	    J^{-1}
	    &\
	    0
	    \\
	    0
	    &\
	    -J^{-1}
	  \end{array}
	  \right)
      \end{array}
      \right)
\end{equation}
whose determinant is given by
\begin{equation}
  \big|U[J-1/2]\big| = (-1)^{N_{f}}J^{-2N_{f}}\,.
\end{equation}
Using this matrix, $|D_{f}[J]|$ in \eqref{deltaf1} decomposes as
\begin{equation}
\label{deltaf2}
  \big|D_{f}[J-1/2]\big| = \frac{1}{|U|}\big|UD_{f}\big| 
    = \frac{1}{|U|}
    \left|
    \begin{array}{cc}
      D_{f}^{\prime}
      &\
      0
      \\
      U^{\prime}
      &\
      \mathbbm{1}
    \end{array}
    \right|
    = (-1)^{N_{f}}J^{2N_{f}}|D_{f}^{\prime}[J-1/2]|
\end{equation}
where $D_{f}^{\prime}[J-1/2]$ is given by
\begin{equation}
\begin{aligned}
  D_{f}^{\prime}[J-1/2]&= \mathbbm{1}\otimes\left(
		      \begin{array}{cc}
			-\frac{J}{2}
			&\
			-\frac{i}{2}
			\\
			-\frac{i}{2}
			&\
			\frac{J}{2}
		      \end{array}
		      \right)
		      -\frac{1}{4J}
		      \left(
		      \begin{array}{cc}
			M^{\dagger}+M^{T}
			&\
			M^{\dagger}-M^{T}
			\\
			M^{\dagger}-M^{T}
			&\
			M^{\dagger}+M^{T}
		      \end{array}
		      \right)
		      \left(
		      \begin{array}{cc}
			M^{\ast}+M
			&\
			M^{\ast}-M
			\\
			M-M^{\ast}
			&\
			-M^{\ast}-M
		      \end{array}
		      \right)
		    \\[5pt]
		    &=
		    \left(
		      \begin{array}{cc}
			-\frac{J}{2}\mathbbm{1} - \frac{r^{2}M^{\dagger}M}{J}
			&\
			-\frac{i}{2}\mathbbm{1}
			\\
			-\frac{i}{2}\mathbbm{1}
			&\
			\frac{J}{2}\mathbbm{1} + \frac{r^{2}M^{\dagger}M}{J}
		      \end{array}
		      \right)
\end{aligned}
\end{equation}
and it's determinant is given by
\begin{equation}
  \big| D_{f}^{\prime}[J-1/2]\big| = \left(\frac{-1}{(2J)^{2}}\right)^{N_{c}} \det\left[\left(J(J+1)+2M^{\dagger}M\right)\left(J(J-1)+2M^{\dagger}M\right) \right]\,.
\end{equation}
Using this result, substituting \eqref{deltaf2} in \eqref{deltaf1} yields
\begin{equation}
\label{1lf}
  \Delta_{f} = \prod_{J=1}^{\infty}2^{-4JN_{c}}\ \prod_{J=1}^{\infty} J^{4J(N_{f}-N_{c})} \prod_{J=0}^{\infty} 
    \det\left(J(J+1)+2M^{\dagger}M \right)^{4J+2}\,.
\end{equation}
Combining \eqref{1lb} and \eqref{1lf}, the one-loop determinant is given by
\begin{equation}
  \Delta = \det (M^{\dagger}M)
\end{equation}
up to an irrelevant divergent factor which may be regularized via zeta function regularization to $2^{2N_{c}/3}$.

\clearpage

\bibliography{refs}
\end{document}